\documentclass{IEEEtran4PSCC}
\pdfoutput=1

\ifCLASSINFOpdf
   \usepackage[pdftex]{graphicx,graphics}
\else
   \usepackage[dvips]{graphicx,graphics}
\fi
\usepackage[cmex10]{amsmath}
\usepackage{amsmath, amssymb}
\usepackage{color}

\usepackage{multirow}
\usepackage{subfig}

\usepackage[justification=centering]{caption} 

\makeatletter
\let\old@ps@headings\ps@headings
\let\old@ps@IEEEtitlepagestyle\ps@IEEEtitlepagestyle
\def\psccfooter#1{%
    \def\ps@headings{%
        \old@ps@headings%
        \def\@oddfoot{\strut\hfill#1\hfill\strut}%
        \def\@evenfoot{\strut\hfill#1\hfill\strut}%
    }%
    \def\ps@IEEEtitlepagestyle{%
        \old@ps@IEEEtitlepagestyle%
        \def\@oddfoot{\strut\hfill#1\hfill\strut}%
        \def\@evenfoot{\strut\hfill#1\hfill\strut}%
    }%
    \ps@headings%
}
\makeatother

\psccfooter{%
        \parbox{\textwidth}{\hrulefill \\ \small{21st Power Systems Computation Conference} \hfill \begin{minipage}{0.2\textwidth}\centering \vspace*{4pt} \includegraphics[scale=0.06]{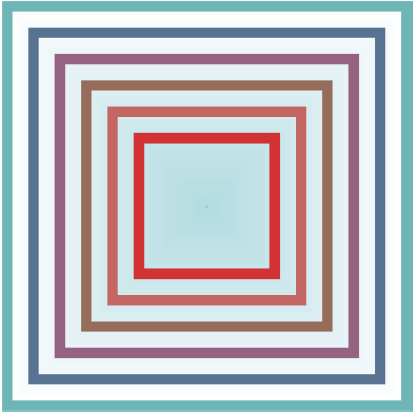}\\\small{PSCC 2020} \end{minipage} \hfill \small{Porto, Portugal --- June 29 -- July 3, 2020}}%
}

\begin{document}
%

\title{Effect of Voltage Source Converters with Electrochemical Storage Systems on \\Dynamics of Reduced-inertia Bulk Power Grids}

\author{
	\IEEEauthorblockN{Yihui Zuo, Mario Paolone}
	\IEEEauthorblockA{Distributed Electrical System Laboratory\\
		\'Ecole Polytechnique F\'ed\'erale de Lausanne\\
		Lausanne, Switzerland\\
		\{yihui.zuo, mario.paolone\}@epfl.ch} 
	\and
	\IEEEauthorblockN{Fabrizio Sossan}
	\IEEEauthorblockA{Center for processes, renewable energies and energy systems\\
		MINES ParisTech\\
		Nice, France\\
		\{fabrizio.sossan\}@mines-paristech.fr}
}


\maketitle

\begin{abstract}
A major concern associated to the massive connection of distributed energy resources is the increasing share of power electronic interfaces resulting in the global inertia reduction of power systems. The recent literature advocated the use of voltage source converter (VSC) interfaced battery energy storage system (BESS) as a potential way to counterbalance this lack of inertia. However, the impact of VSCs on the dynamics of reduced-inertia grids is not well understood especially with respect to large transmission grids interfacing a mix of rotating machines and resources interfaced with power electronics. 
In this regard, we propose an extension of the IEEE 39-bus test network to quantify the impact of VSCs on reduced-inertia grids. In this respect, a reduced-inertia 39-bus system is obtained by replacing 4 synchronous generators in the original 10-synchronous machine system, with 4 wind power plants modeled as aggregated type-3 wind turbines. Then, a large-scale BESS is integrated into the reduced-inertia network via a three-level neutral-point clamped (NPC) converter, thereby to be used for studying the impact of VSC on the dynamics of the inertia-reduced power system, as well as for comparing different VSC controls. The proposed models are implemented on a real-time simulator to conduct post-contingency analysis, respectively, for the original power system and the reduced-inertia one, with and without the BESS-VSC. 
\end{abstract}


\thanksto{\noindent This work is part of the OSMOSE project. The project has received funding from the European Union’s Horizon 2020 research and innovation programme under grant agreement No 773406. This article reflects only the authors views and the European Commission is not responsible for any use that may be made of the information it contains.}

\section{Introduction}
Modern power systems are characterized by large shares of resources interfaced with power electronics. 
In European Union, the renewable energy shares vary from 5\% to 54\%, while many countries encounter penetration levels of renewable generation (i.e.,wind and solar) in excess of 15\% of their overall annual electricity consumption~\cite{IRENAeurop}. Some power systems (e.g. in Spain, Portugal, Ireland, Germany and Denmark) have even already experienced instantaneous penetration levels of more than 50\% of converter connected generation~\cite{EU_ReP2013}.
As generally acknowledged, the large deployment of non-synchronous generation will determine a reduction of the system inertia and thus lead to very fast dynamics in case of contingencies, as indicated in several TSO reports~\cite{AEMO,RG-CEreport,ERCOTreport}. 
An example is the severe blackout happened in the South Australian power system in 2016, when a wind storm hit the region while half of the power consumption was fed by wind generation~\cite{AEMO}, causing the grid frequency to decrease with a rate of change of 6.25 Hz/s.
In this context, fast-ramping devices, such as converter-interfaced sources, may provide fast primary control response and are regarded as a potential and advocated remedy for power grid frequency regulation~\cite{RG-CEreportFRS}.

To address the challenges related to reduce levels of system inertia, battery energy storage systems (BESSs) are broadly advocated as one of the potential solutions~\cite{achdispatch,lowinertiaRa} thanks to their large ramping rates capacities. 
Utility-scale BESSs, which are now commercially available, are also recognized for other desirable features, including high-round-trip efficiency, and long cycle-life~\cite{SciBattery4Grid}. 
BESSs are interfaced to the public AC power grid through four-quadrant voltage converters~\cite{reveiwbessconverter}, which can be typically controlled at a sub-second resolution and used to provide grid ancillary services ranging from fast primary frequency response up to energy management (possibly, multiple \cite{emilbessmultservice}).

There are generally two main approaches to achieve the power control for power converter-interfaced units: grid-following control and grid-forming control~\cite{ref1_defineVSCcontrol,ref2_defineVSCcontrol,PSCC_PWRS_survey}.
A grid-following unit is based on a power converter injecting required active and reactive power via modifying the amplitude and angle (with respect to the grid voltage phasor) of the converter reference current, with the requirement on the knowledge of the fundamental phasor of the grid voltage at a point of common coupling (PCC). A grid-forming unit is based on a voltage source converter (VSC) that controls the frequency and voltage at a PCC, behaving as a voltage source behind an impedance and without requiring the knowledge of the fundamental frequency phasor of the grid voltage at the PCC. In case a grid-forming control is used to regulate the converter injected power, the knowledge of the grid voltage phasor is required. In this respect, the concept of grid-supporting mode was introduced in~\cite{ref2_defineVSCcontrol} where additional high level control loops are incorporated into the grid-forming and grid-following control, to regulate the AC voltage via power output.

On one hand, in actual power system, the majority of converter-interfaced resources is controlled as grid-following sources as this operation mode is considered to be efficient for the load resources~\cite{CurrentgrieCSC1,CurrentgrieCSC2}. As mentioned above, it relies on the knowledge of fundamental phasor of the grid voltage at the PCC, which can be tracked via a Phase Locked Loop (PLL). On the other hand, the future low-inertia grid may require large amount of grid-forming devices that provide a specific support for frequency and voltage regulation and stability, black-start capabilities, as well as  synchronization mechanisms~\cite{PSCC_PWRS_survey,challengeoflowinertagrid}.



{\color{red}}
To the authors' best knowledge, very few researches have attempted to quantitatively assess the effects of inertia reduction and deployment of grid-scale VSC-based BESS on the dynamics of bulk power systems by including detailed dynamic models of the grid and its components. The work in \cite{dyn4lowinertia1} uses detailed models of two multi-area systems, providing insights on their dynamic behaviors when subject to large installed capacities of wind generation.
In~\cite{bess_f_ROCOF}, the inertia of the IEEE 39-bus system is tailored to resemble the relative low-inertia Irish system; then, the ameliorating impact of a BESS, implemented as a negative load injection, on grid frequency transients is investigated.
Even if the works in ~\cite{dyn4lowinertia1,bess_f_ROCOF} use detailed dynamic simulation models of the grid, they adopt a simple model for the power converter-interfaced units, thus failing in capturing and assessing the interactions between VSC-based resources and the grid.

In this context, the paper proposed a study based on the detailed dynamic models of grids, converters and controls to analyse the impact of inertia reduction on power systems and the influence of VSC control approaches (grid-following versus grid-forming) on the dynamics of reduced-inertia power systems.
To this end, starting from the IEEE 39-bus benchmark system, we derive two new system configurations that allow us to evaluate the system behavior in a reduced-inertia setting while considering VSC-based BESS:
\begin{itemize}
	\item A reduced-inertia 39-bus power system, created by replacing 4 synchronous generators with 4 aggregated type-3 wind power plants;
	\item A reduced-inertia 39-bus power system, created by replacing 4 synchronous generators with 4 aggregated type-3 wind power plants and introducing a VSC-based BESS.
\end{itemize}

The paper is structured as follows: Section~\ref{reduceinertia39} introduces the dynamic simulation models for the reduced-inertia 39-bus power grids, Section~III describes the dynamic models for the VSC-based BESS associated with a PLL-free grid-forming control and grid-following control, and Section~IV presents and discusses the simulation results. Finally, Section~V summarizes the results and provides indications of the control laws to be used for VSCs connected to limit the potential problems associated to reduced-inertia power systems.

\begin{figure}[t]
	\centering
	\includegraphics[width=0.75\linewidth]{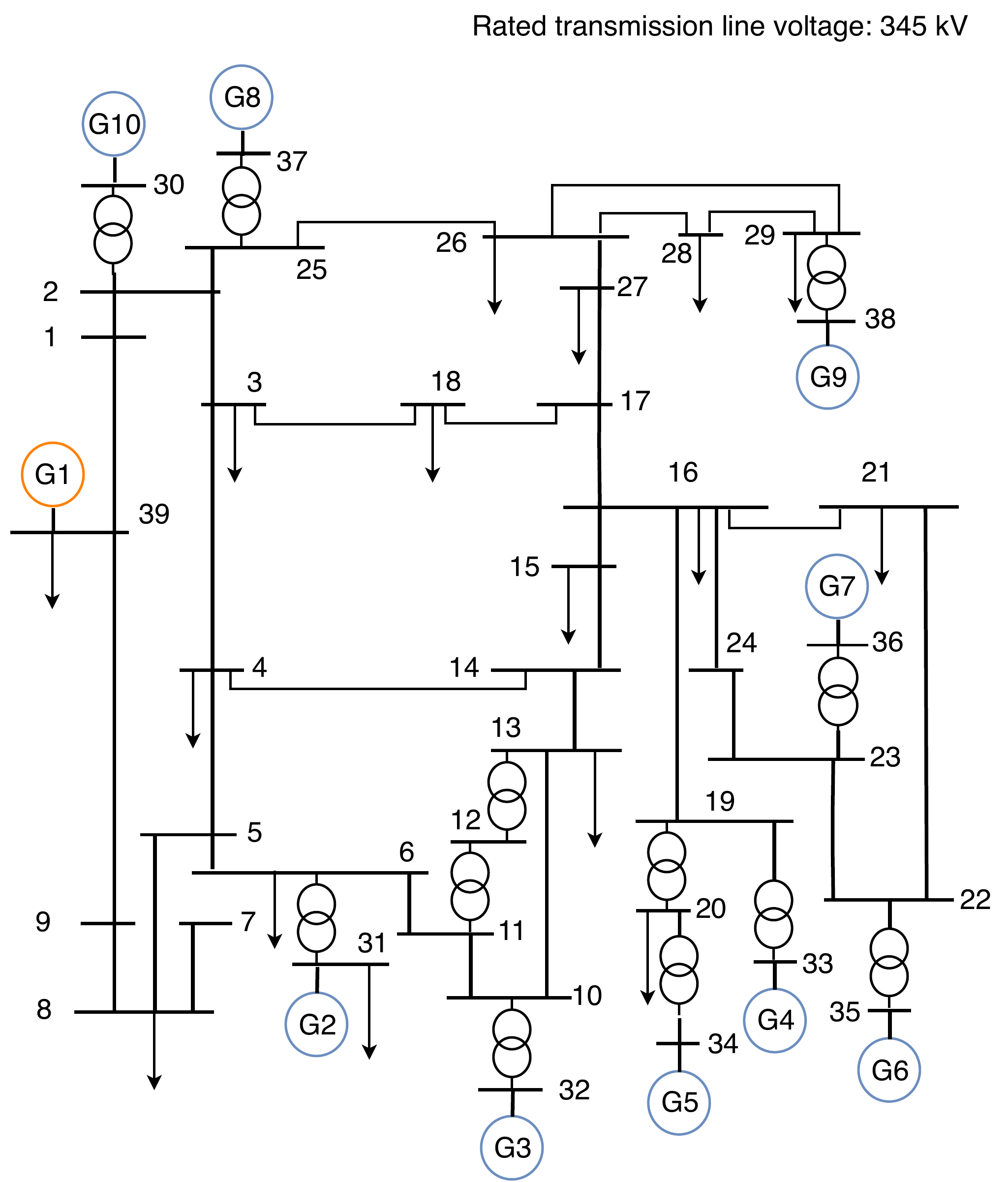}
	\caption{Topology of IEEE 39-bus benchmark test network.}
	\label{fig:39_con1}
\end{figure}

\begin{figure}[t]
	\centering
	\includegraphics[width=0.75\linewidth]{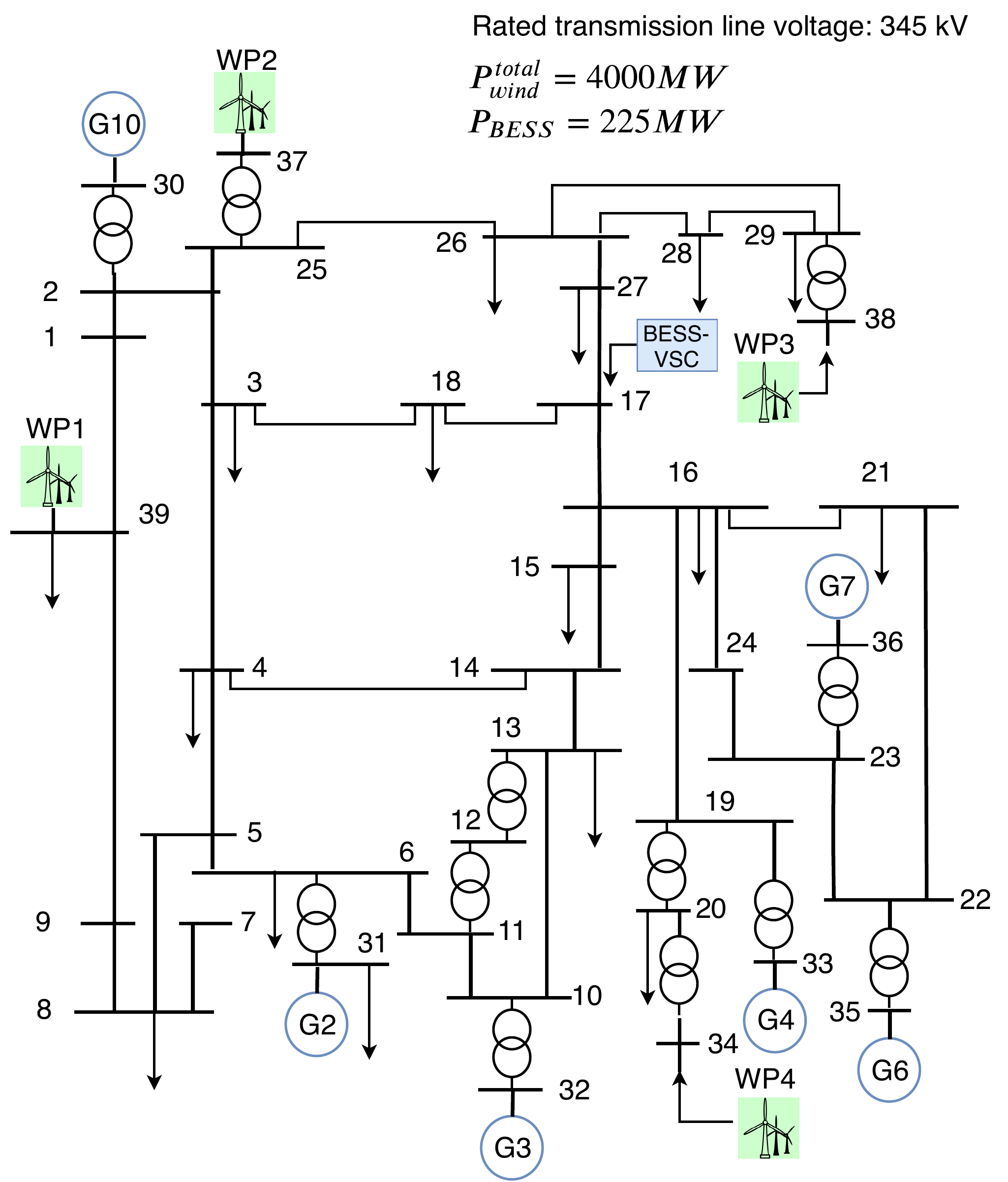}
	\caption{Topology of reduced-inertia 39-bus power system (The presence of the BESS at bus 17 is taken into account in the Config.~II).}
	\label{fig:39_con2}
\end{figure}

\section{Reduced-inertia bulk power system}\label{reduceinertia39}

All dynamic models presented in this and the next section are built in MATLAB/Simulink and executed in an OPAL-RT eMEGAsim real-time simulator. For the sake of reproduciblilty, all the proposed models are open-source and freely available online~\cite{github-39busgrids}, where the modeling details and parameters used in the proposed models are all provided.

The IEEE 39-bus benchmark test network, shown in Fig.~\ref{fig:39_con1}, has been widely adopted for studies of power system dynamics since it first appeared in~\cite{EPRIorignal39_1}. We modified the IEEE 39-bus benchmark power system by replacing 4 synchronous generators (denoted in Fig.~\ref{fig:39_con1} as G1, G5, G8 and G9) with 4 wind power plants based on an aggregated model of a type-3 double-fed induction generator (DFIG) wind turbine, as shown in Fig.~\ref{fig:39_con2}.
This allows us to model a scenario with reduced system inertia due to displacing a part of conventional synchronous generation capacity in favor of converter-interfaced production. Table.~\ref{configIvsII} reports the total value of the inertia constant (referred to a 100 MW base and obtained by summing the inertia constant of the all conventional power plants) for the original grid and the modified grid, which are referred to as Config.~I and  Config.~II, respectively.

Correspondingly, we create two full-replica dynamic models for Config.~I and Config.~II. This modelling details are provided in the followings of this section.

\begin{table}[t]
	\renewcommand{\arraystretch}{1.3}
	\centering
	\caption{Inertia constant for Config.~I and Config.~II.}
	\label{configIvsII}
	\begin{tabular}{|c|c|c|}
		\hline
		 & Config.~I &Config.~II\\
		\hline
		H [s] & 784.7 & 197.9\\
		\hline
	\end{tabular}
\end{table}



\subsection{Synchronous generators}
Conventional generation consists of hydro- and thermal-power plants. They are simulated with of a sixth-order state-space model for the synchronous machine, a prime mover \cite{report1973dyn}, a DC1A excitation system associated with an AVR~\cite{stdp421.5-d38-2015}. 
The generator model includes the primary frequency regulation with a static droop coefficient $R_p=5\%$. The power plant G7 also implements a secondary frequency regulator with an integration time constant of 120~s.

\subsubsection{Synchronous machine}The generator model provided in the original technical report~\cite{EPRIorignal39_1} is essentially a fourth-order generator model, as it does not include the subtransient circuits. Therefore, we use a sixth-order state-space model for the synchronous machine, whose synchronous and transient parameters 
are taken from the original technical
report~\cite{EPRIorignal39_1}, while the subtransient parameters 
are inspired from real-world test parameters, adapted from the IEEE Std. 1110™-2002(R2007)~\cite{IeeeguideforSG} and in the EPRI technical reports \cite{EPRISG1,EPRISG2}. 

\subsubsection{Hydraulic turbine and governor system} We adopt the commonly-used standard hydro turbine governor model as illustrated in~\cite{HTgovnorPES}. 
According to~\cite{IEEETurGov2011}, the response of the turbine governing system should be tuned to match the rotating inertia, the water column inertia, the turbine control servomotor timing and the characteristics of the connected electrical load. Therefore, as recommended in~\cite{IEEETurGov2011}, we use $T_{M}=2H$ and $T_{M}:T_{w}=3:1$.
 $H$ is the generator inertia constant, $T_M$ is the  mechanical inertia constant, and $T_w$ is water inertia time (also known as "water starting time"). 
	The PI governor parameters are derived according to~\cite{HydraTut1992}, where $1/K_{P}= 0.625T_{W}/H$ and $K_{P}/K_{I} = 3.33 T_{W}$.

\subsubsection{Steam turbine and governor system} The steam turbine and governor model are adapted from~\cite{report1973dyn}, where the steam turbine system is presented as tandem-compound, single mass model and the speed governor consists of a proportional regulator, a speed delay and a servo motor. The parameters for the steam turbine-governor are taken from the typical values used, for instance, in~\cite{report1973dyn,kundur1994power}.
\subsubsection{Excitation system} The IEEE DC type 1 exciter associated with an AVR~\cite{stdp421.5-d38-2015} is implemented in the excitation system, whose parameters are adapted from~\cite{ieee2015powerTR18}.

\subsection{Dynamic loads}
In order to reproduce a plausible dynamic load behavior, the EPRI LOADSYN model has been adopted~\cite{loadrepresentation}.
Specifically, we implemented the three-phase dynamic and voltage-dependent load model based on the following equations:
\begin{align}
    P(t) = P_{0}(t)\left(\frac{V(t)}{V_{0}}\right)^{K_{pv}}[1+K_{pf}(f(t)-f_{0})] \\ 
   Q(t) = Q_{0}(t)\left(\frac{V(t)}{V_{0}}\right)^{K_{qv}}[1+K_{qf}(f(t)-f_{0})]
\end{align}

where $P(t)$ and $Q(t)$ are the three-phase load active and reactive power. The coefficients $K_{pv} $, $K_{pf}$, $K_{qv}$, $K_{qf}$ are obtained from typical load voltage and frequency parameters inferred from EPRI LOADSYN program. 
In this regard, we represent $f(t)$, $V(t)$, $P_0(t)$, and $Q_0(t)$ as time-varying variables sampled with a resolution of 20~ms. We assume that $P_{0}(t)$ and $Q_{0}(t)$ are active and reactive power consumed at rated frequency and voltage. The rated demand profile is adapted from a monitoring system based on PMUs installed on the 125~kV sub-transmission system of Lausanne, Switzerland~\cite{Derviskadic-etAl2018-tsg}. Coherently with the other model variables, the measured time-series power data are sampled with a resolution of 20~ms. Since the nominal load values in the original IEEE 39-bus power system are different from our measured data, the final demand patterns are obtained by re-scaling the measured time series with respect to the rated power in~\cite{EPRIorignal39_1}. 
\begin{figure}[t]
	\centering
	\includegraphics[width=0.75\columnwidth]{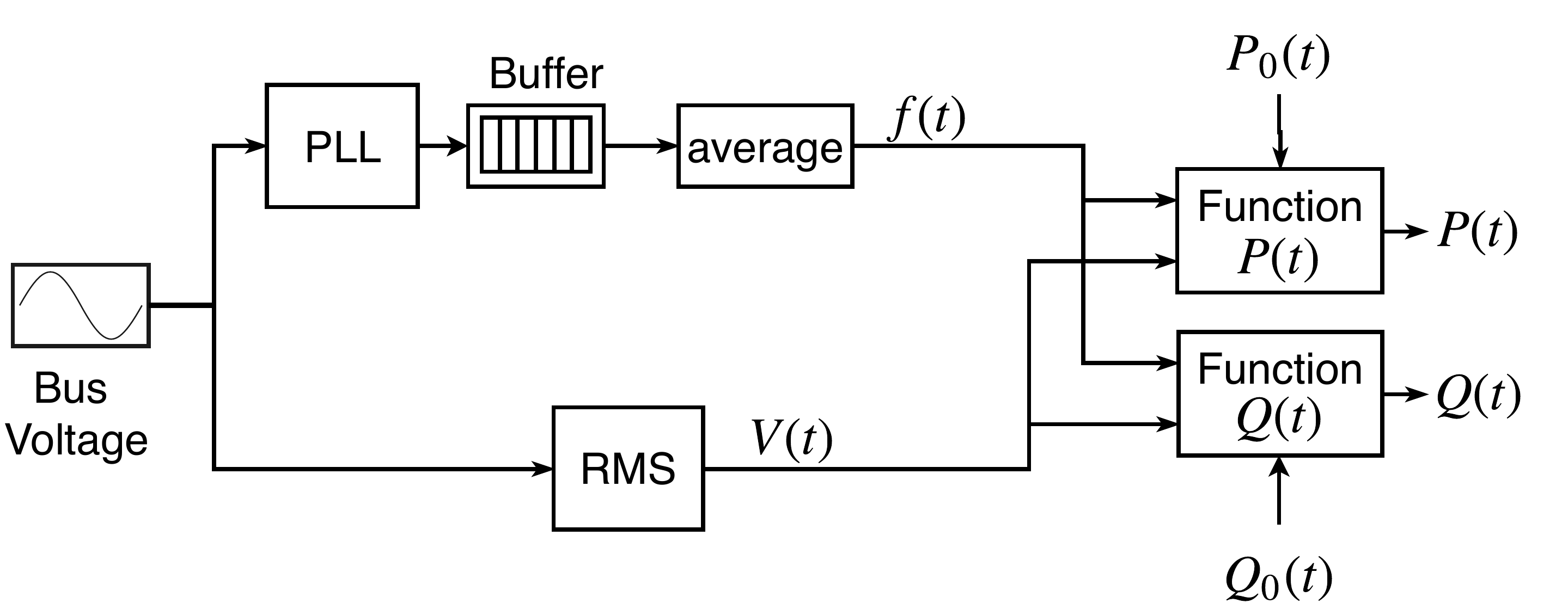}
	\caption{Diagram of the EPRI LOADSYN dynamic load model.}
	\label{fig:dynamicload}
\end{figure}

The  implementation of the EPRI LOADSYN model is illustrated in Fig.~\ref{fig:dynamicload}. A conventional Phase Locked Loop (PLL) and a Root Mean Square (RMS) operator measure the bus frequency and voltage to be employed in the dynamic load model. 
On one side, as the PLL may be inaccurate in transient conditions, a moving average mechanism is implemented in order to avoid improper behavior of the dynamic load model. Specifically, the PLL-tracked frequency is updated every 1~ms, and then buffered for averaging. {\color{red}}
The overall buffer size is 240 samples, with an overlap size of 220 samples (i.e., the final frequency $f(t)$ is reported every 20~ms). 
On the other side, the bus voltage $V(t)$ is given by a RMS operator that computs over a window length of 240~ms and  reports every 20~ms.

\subsection{Wind power plants}
	\begin{figure}[t]
		\centering
		\includegraphics[width=1\linewidth]{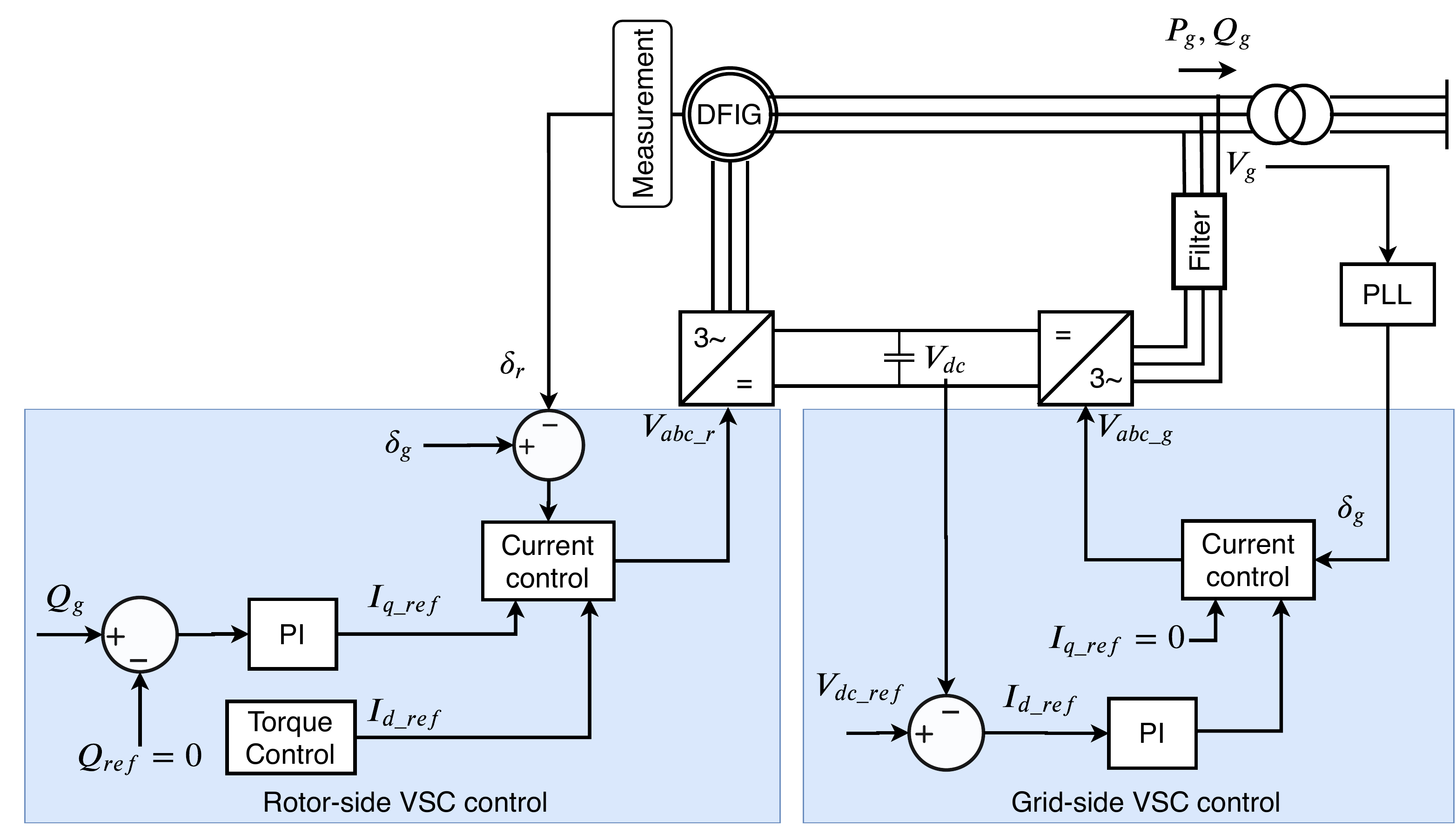}
		\caption{Diagram of wind power plant and controls.}
		\label{fig:diagram of wind plant} 
	\end{figure}
	
The wind power plants are modeled as proposed in~\cite{windfarm}. In particular, the power output is approximated by multiplying the power output of a detailed model of a single wind turbine to match the total nominal capacity of the whole wind farm. The diagram of the overall system in shown in Fig.~\ref{fig:diagram of wind plant}. 
Each wind generator model consists of a DFIG and an averaged back-to-back converter model~\cite{windgen}. For this analysis, the detailed aerodynamic model of wind turbine is not involved, as its effect is accounted already in the wind profiles. 
The wind power profiles are generated at 1 second resolution by re-sampling the measurements at 1 minute resolution from~\cite{energy2009analysis}.
The re-sampling approach is based on the statistical characteristics of the aggregated wind generation profiles presented in~\cite{coughlin2010analysis}.
More details about producing wind power profiles are described in~\cite{DispatchandPFR}.
	
The back-to-back VSCs are modelled as equivalent voltage sources. In this average converter model, the dynamics resulting from the interaction between the control system and the power system are preserved. As shown in Fig.~\ref{fig:diagram of wind plant}, two grid-following controls are implemented in the back-to-back converters. The rotor-side converter controls active and reactive power through rotor current regulation whilist the stator-side converter regulates DC bus voltage and permits operation at a constant power factor (i.e., zero reactive power). 
	
		

\section{Voltage source converter interfaced battery energy storage system }\label{BESS-VSC}

We install a detailed model of a BESS at bus 17 in the reduced-inertia 39-bus power system. As detailed below, it consists of the battery cell stack (necessary to model voltage dynamics on the converter DC bus), and the power converter, which is modelled at the level of the switching devices. 
\begin{figure}[t]
\centering
\includegraphics[width=0.6\columnwidth]{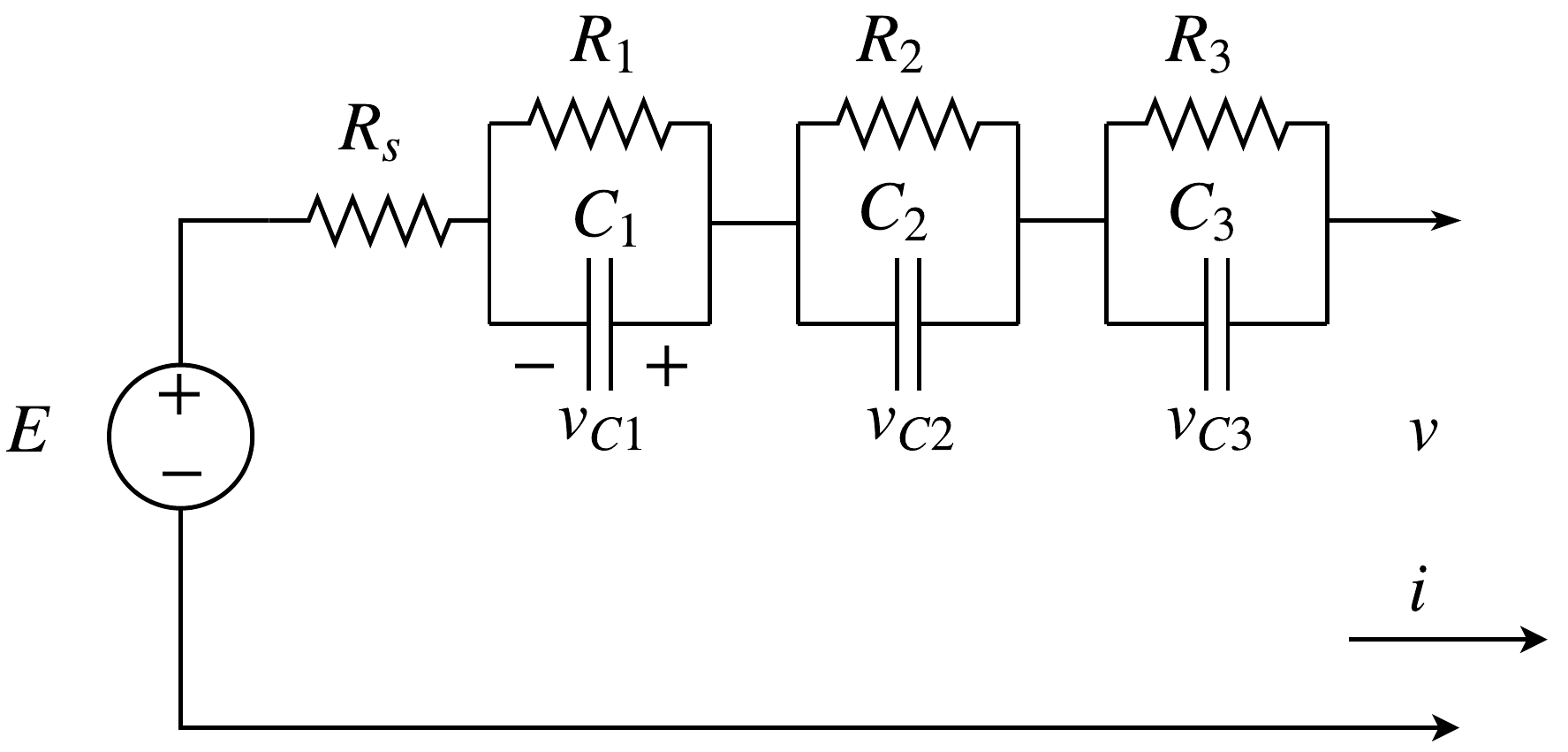}
\caption{Three-time constant equivalent circuit of the battery cell stack.}
\label{fig:TTC}
\end{figure}

\begin{figure}[t]
	\centering
	\includegraphics[width=1\columnwidth]{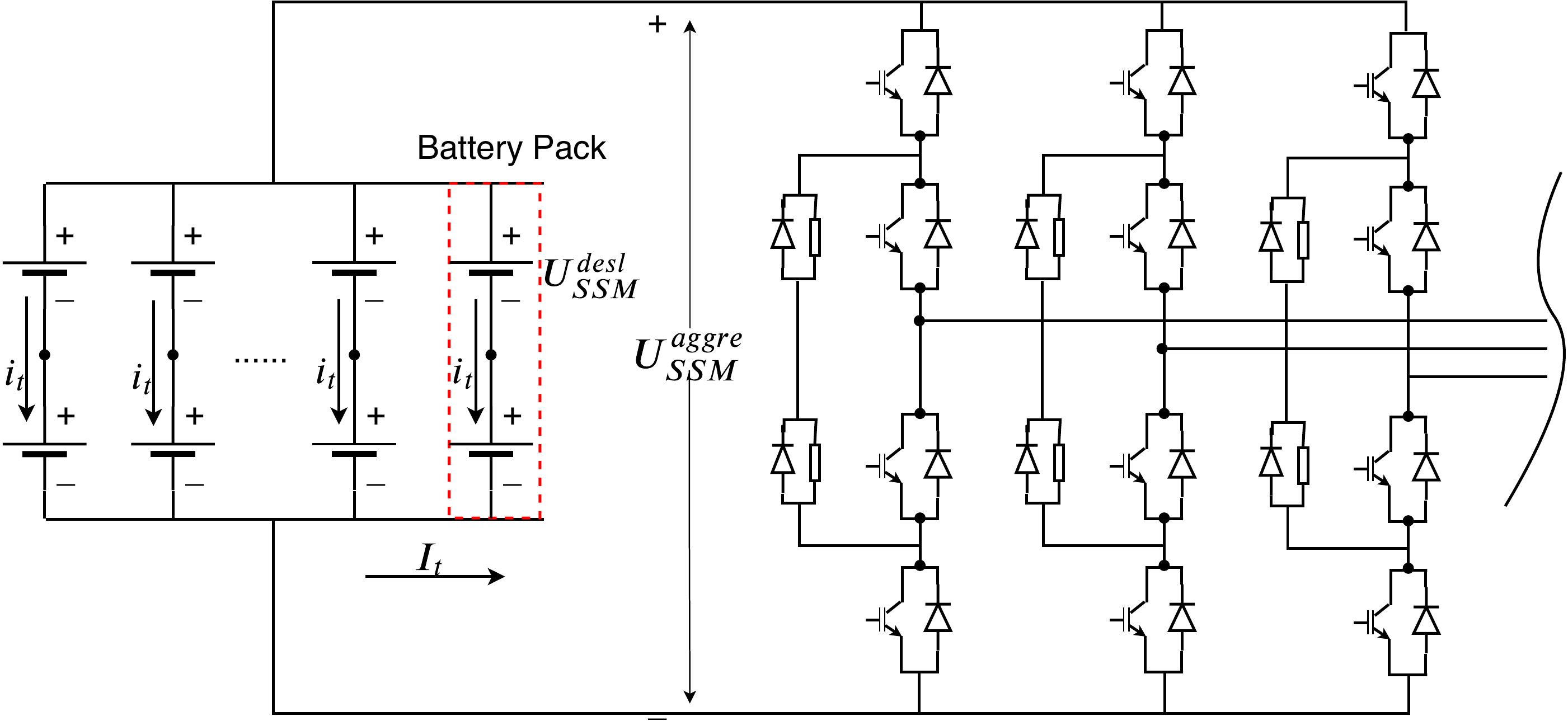}
	\caption{Diagram of BESS-VSC.}
	\label{fig:bessvsc}
\end{figure}


\subsection{Battery cell stack}

The voltage at the terminal of a battery is generally dynamic and it depends on the output current, state-of-charge, cells temperature, ageing conditions, and C-rate. In control applications, it is typically modelled with electric equivalent circuits, which trade detailed modelling of the electrochemical reactions for increased tractability, see e.g. \cite{1634598, YannLiaw2004835}.
\begin{table}[!t]
	\centering
	\renewcommand{\arraystretch}{1.1}
	\caption{Parameters of BESS to be connected to HV transmission grid}
	\label{rtsocttc}
	\begin{tabular}{|c|c|c|c|c|c|}
		\hline 
		\textbf{SOC [\%]} & \textbf{0-20 }& \textbf{20-40} & \textbf{40-60} & \textbf{60-80} & \textbf{80-100} \\ 
		\hline 
		$E$ [$V$] &1184.4 &1250.0 &1305.8&1360.4&1466.4 \\ 
		$R_{s}$ [$\Omega$] & 0.052 & 0.042& 0.030  & 0.028  & 0.026 \\ 
		$R_{1}$ [$\Omega$] & 0.190  & 0.150 & 0.180 & 0.158 & 0.398 \\ 
		$C_{1}$ [$F$]& 4465 &4904.5&6998 &6000 &5617\\ 
		$R_{2}$ [$\Omega$]& 0.08 & 0.018 & 0.018 & 0.018 &  0.020\\ 
		$C_{2}$ [$F$] &454.5  &1069.5 &1241 &1245 &1252.5\\
		$R_{3}$ [$\Omega$] & 5.0e-3  & 9.8e-5  &4.8e-4  & 13.6e-4 &  12.0e-4 \\ 
		$C_{3}$ [$F$]&272.1 &394.5&1479.8&2250&3088.7\\ 
		\hline 
	\end{tabular} 
\end{table}
In this paper, we use a validated grey-box model identified from measurements of a 720 kVA/560 kWh Lithium-titanate-oxide battery at EPFL \cite{achdispatch}. The model is a third-order model with parameters that depend on the state-of-charge. Despite most of literature refers to two-time-constant models (i.e., second order models), it was shown in \cite{achdispatch} that when considering voltage measurements at a second resolution, a third state is necessary to explain system dynamics. The three-time constant equivalent circuit of the battery cell stack is shown in Fig.~\ref{fig:TTC}. The state-space representation of the model is:
\begin{align}
\dot{x}(t)=Ax(t)+Bu(t) \\
y(t)=Cx(t)+Du(t) \label{eq:statespace}
\end{align}
where
\begin{align}
& A = \text{diag}\left(-1/(R_1C_1), -1/(R_2C_2), -1/(R_3C_3))\right) \\
& B = \begin{bmatrix}
1/C_1 & 0\\
1/C_2 & 0\\
1/C_3 & 0
\end{bmatrix},
C= \begin{bmatrix}
1 &1 &1
\end{bmatrix},
D= \begin{bmatrix}
R_{s} & E
\end{bmatrix} \\
&x=
\begin{bmatrix}
v_{C1} & v_{C2} & v_{C3}
\end{bmatrix}, ~
u(t)=
\begin{bmatrix}
{i_{t}/156} & 1
\end{bmatrix}^T. \label{eq:statespace_f}
\end{align}
Model output $y(t)$ denotes the terminal voltage, and input $i_t$ is the total DC current absorbed/provided by the battery. The elements of matrices $A,~B,\text{and }D$ are state-of-charge-dependent and can be identified from measurements, as described in \cite{achdispatch,namor_isgt_2018}.  However, since the power rating of the BESS that we use in this work (225 MVA) is larger than the one for which the model is proposed in \cite{achdispatch} (0.72 MVA), we need to adapt the model parameters as described in the following. First, we achieve the target power (225 MVA) with a configuration composed of two cell stacks in series and 156 in parallel. The two units in series are explained by the fact that, in the attempt of increasing the voltage on the DC bus (to reduce losses), this is the largest (integer) number of series elements that a converter can accommodate given that the original model refers to a battery with an open-circuit voltage of 800 V at full charge and power electronic can conveniently handle voltage up to 2 kV. By assuming that all the paralleled battery packs are identical, the voltage of the aggregated BESS is considered equal to the voltage of each battery pack. The parameters of the equivalent circuit models are obtained by doubling all the parameters reported in \cite{achdispatch}, except for capacitors, whose values were halved to retain the same time constants as those identified. Final parameters adopted for three-time constant model \eqref{eq:statespace}-\eqref{eq:statespace_f} are reported in in Table~\ref{rtsocttc}. The total BESS current $i_k$
is used to compute the state-of-charge:
\begin{eqnarray}
\text{SOC}_{k+1}=\text{SOC}_{k}+\frac{Ts}{3600}\frac{i_{k}}{C_{nom}}\label{eq:soc}
\end{eqnarray}
where $T_s=0.001$~s is the sampling time and $C_{nom}=$117~kAh is the BESS capacity.

  \begin{figure}[t]
	\centering
	\includegraphics[width=1\columnwidth]{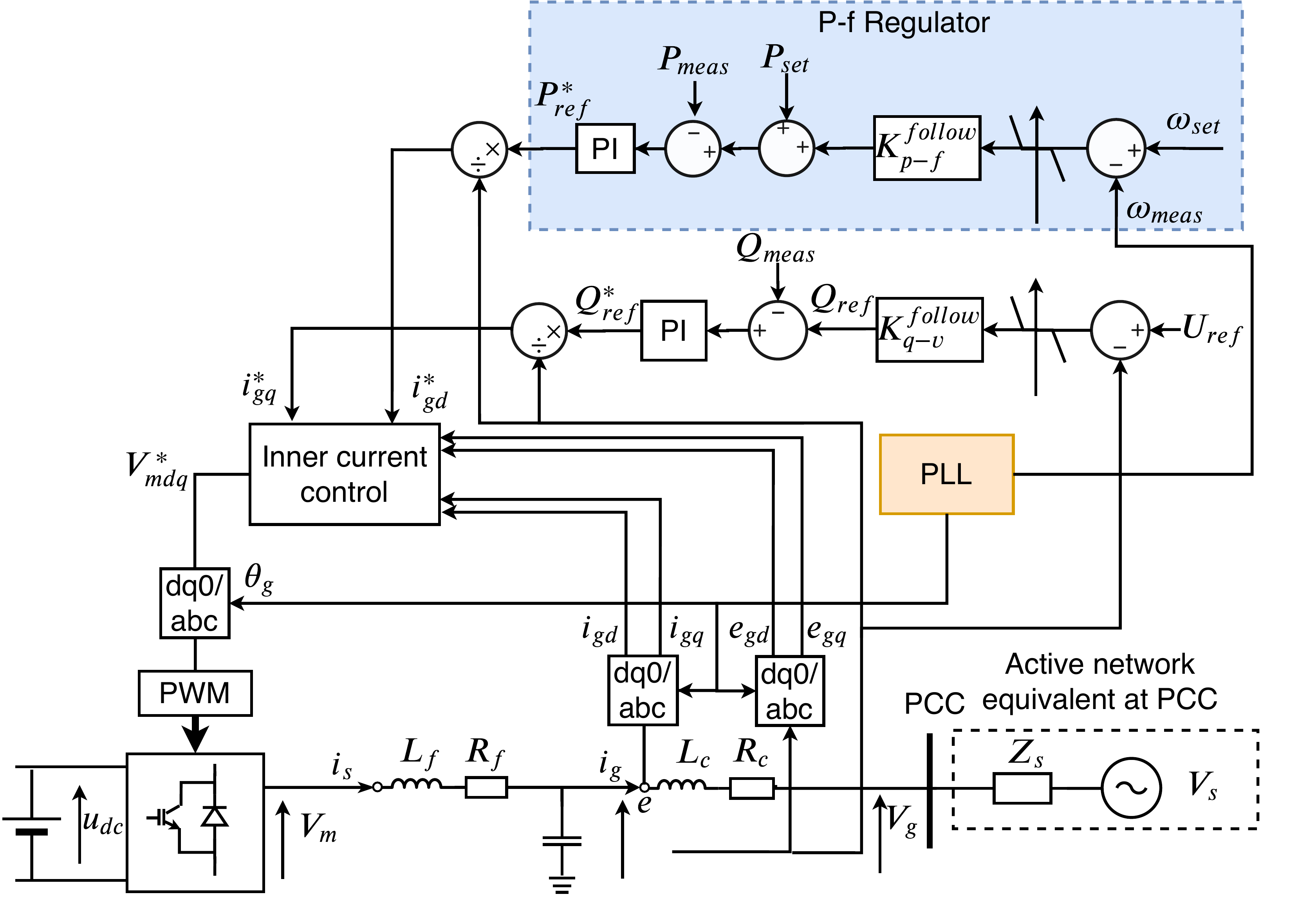}
	\caption{Grid-following converter with grid-supporting mode.}
	\label{fig:gridfollowing}
\end{figure}
  	\begin{figure}[t]
	\centering
	\includegraphics[width=1\columnwidth]{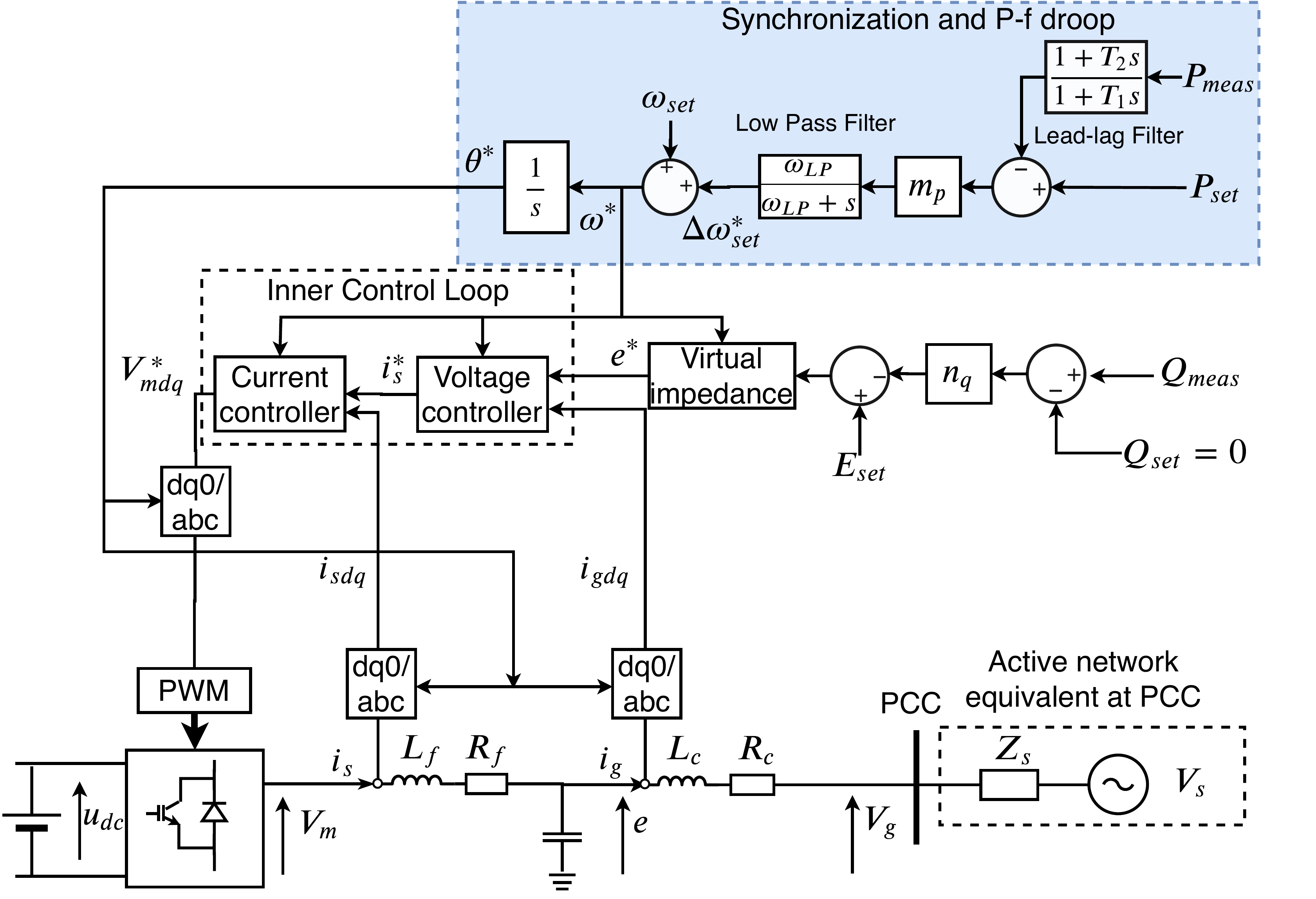}
	\caption{PLL-free grid-forming converter.}
	\label{fig:gridforming}
\end{figure}

  \begin{figure}[!t]
	\centering
	\includegraphics[width=1\columnwidth]{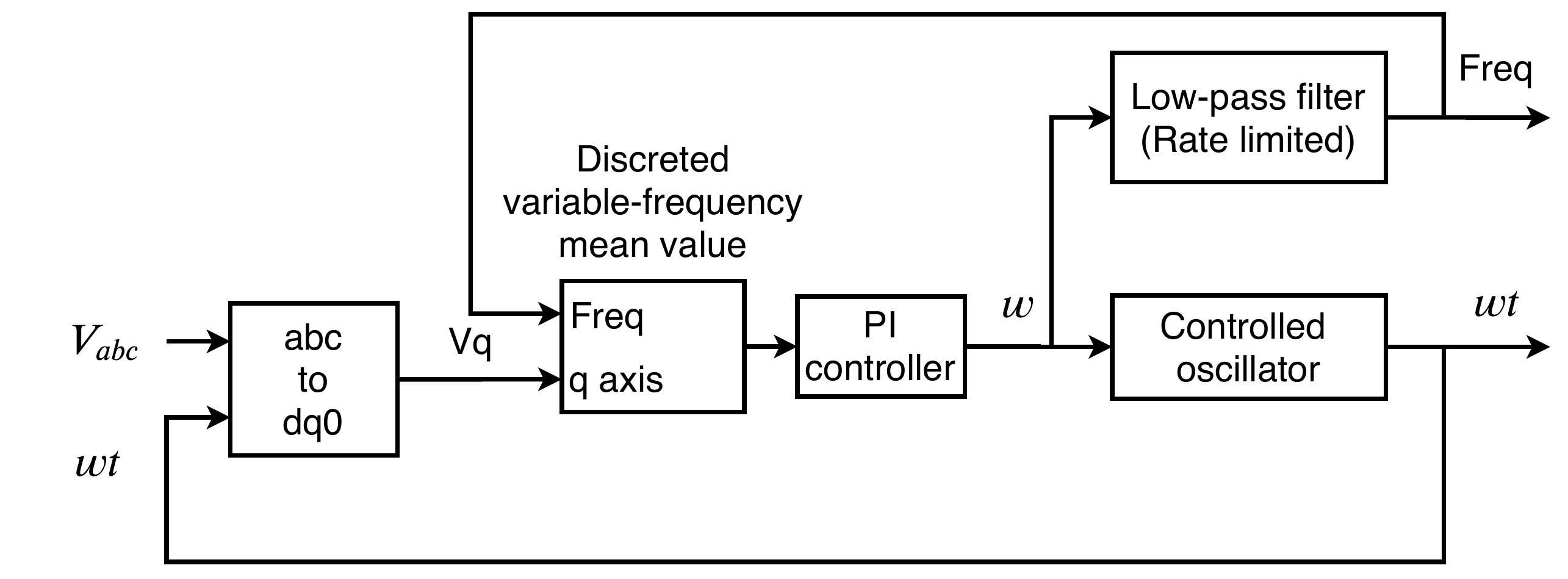}
	\caption{PLL implemented in grid-following converter with grid-supporting mode.}
	\label{fig:gridfollowing_PLL}
\end{figure}

\subsection{Power electronic converter}

The model of the power converter consists of a fully modeled three-level neutral-point clamped (NPC) converter, consisting of 12 IGBT/Diode pairs and 6 clamp diodes. It is shown Fig.~\ref{fig:bessvsc}.
To be applicable for the real-time simulations, the ARTEMiS state-space nodal (SSN) blocks are used to assign the three arms of the converter into three separate SSN groups. This allows the solvers to decouple the large state-space equation into smaller groups~\cite{SSMforRT}.

\subsection{Controls for voltage source converter}

We choose two converter controls, namely the the grid-following control with support mode and the PLL-free grid-forming control, as shown in Fig.~\ref{fig:gridfollowing} and Fig.~\ref{fig:gridfollowing}.

\subsubsection{Grid-following converter operated with grid-supporting mode}
The grid-following control adjusts the injected power with respect to the grid voltage at the PCC, whereas the grid-forming control adjusts the modulated voltage with respect to the grid voltage at PCC. Details of the considered control schemes are described in the followings. 


The grid-following control has been widely deployed in grid-connected converters, such as in VSC-HVDC~\cite{vschvdcreview} and the back-to-back converter of wind power plants (type-3 and type-4 wind turbine generators)~\cite{windvsccontrol}.

As shown in Fig.~\ref{fig:gridfollowing}, the adopted grid-following control injects the required amount of active and reactive power by controlling the injected current with a specific phase displacement in respect to the grid-voltage at a PCC. Therefore a phasor estimation device (i.e., PLL) is required to estimate the fundamental frequency phasor of the grid voltage, so as to generate the instantaneous value of the current reference and eventually the voltage reference. In this regard, the active and reactive power are controlled independently. 

Fig.~\ref{fig:gridfollowing_PLL} shows the three-phase PLL used for tracking the fundamental phasor of grid-voltage at the PCC. The three-phase input signals are converted to the dq0 rotating frame using the angle provided by a controlled oscillator. The q-axis of the signal is filtered with a discrete mean block that computes the mean value of q-axis voltage over a sliding window of one cycle whose frequency is the one of the previous estimation. The PI controller output, corresponding to the angular velocity, is filtered and converted into the frequency. The proportional gain and integral gain for the PI controllers are $K_{p,PLL}$=60 and $K_{I,PLL}$=1400, respectively.

The grid-following converter is operated with grid-supporting mode by adding a high level frequency and voltage regulators with droop characteristics.
The active power is regulated according the droop coefficient $K_{p-f}^{following}=20$, as the difference between the measured frequency (from PLL) and the frequency reference exceeds the dead-band of 0.001 p.u. The reactive power in regulated according the droop coefficient $K_{q-v}^{following}=10$, as the difference between the measured voltage and voltage reference exceeds the dead-band of 0.005 p.u.

 \subsubsection{PLL-free grid-forming converter}
 The grid-forming control allows the converters operating as synchronized voltage source. Thereby, they do not require an explicit current control. As stated in the introduction, they can use the angle difference between the grid voltage and the modulated voltage to control power. In this context, the estimate of grid voltage angle is necessary and can be achieved in two ways: use a PLL to estimate the grid voltage angle or, instead, directly link the active power exchange to the angle difference between the grid voltage ($\theta_g$) and the modulated voltage ($\theta_m$) to create a PLL-free controller. 
 
 We adopt the PLL-free grid-forming control proposed in~\cite{Migratereport} and developed for VSC connecting at transmission level~\cite{ImproveMigrate}.  
 Fig.~\ref{fig:gridforming} shows the control diagram of the adopted PLL-free grid-forming control.
Such a control architecture creates a link between the output voltage angle of the converter and the active power which not only enables the synchronization with the grid but also allows the converter to deliver in primary frequency regulation.

As shown in the blue sub-diagram in Fig.~\ref{fig:gridforming}, the output voltage angle is directly linked with the difference between measured active power and reference active power. 
Specifically, $m_p=0.05$ corresponds to frequency droop coefficient $K_{p-f}^{forming}=20$.
A first-order low-pass filter is added to avoid fast frequency variations and to filter the power measurements noise, and a lead-lag filter is implemented on the power measurement to improve the converter dynamics~\cite{gridformingreviewGuil}. 
According to~\cite{Migratereport}, the cut-off frequency for the low pass filter is $\omega_{LP}=31.4$~rad/s. The adopted time constants for the lead-lag filter are  $T_1=0.0333$~s and $T_2=0.0111$~s.

The considered PLL-free grid-forming control is an effective simple scheme that allows the converter to synchronize with the power grid and to provide the primary frequency regulation services. However, the active and reactive power are not decoupled because the reactive power is coupled with the angle difference between grid voltage and the modulated voltage ($\delta=\theta_g-\theta_m$). In particular, we have that (see Fig.~\ref{fig:gridforming}):
\begin{eqnarray}
     Q = \frac{V_g}{R_C^2+X_C^2}  [R_{C}V_{m}sin(\delta)+X_{C}(V_{g}-V_{m}cos(\delta))]
\end{eqnarray}
where the modulated voltage angle $\theta_m$ is determined by the active power control, $V_m$ is the converter AC voltage amplitude, and $V_g$ is the amplitude of the grid voltage at PCC. $R_C$ and $X_C$ are the transformer impedance components as shown in Fig.~\ref{fig:gridforming}.



\section{Dynamic simulations}
\subsection{Impact of Inertia Reduction}\label{ConfigIvsConfigIIresult}
To evaluate the systems response in extreme condition without the presence of converter-interfaced BESS, we reproduce a contingency (i.e., the tripping of generator G6) in both Config.~I and Config.~II. Table~\ref{innodal} reports the initial nodal power injections\footnote{The reactive power provided by the wind power plants are generated by shunt capacitors.} (i.e. pre-contingency power injections) for Config.~I and Config.~II. It shows that, in Config.~II, wind generation accounts for more than half of the total active power injection, i.e., 3789~MW versus 7129~MW. 

\begin{figure}[t]
	\centering
	\includegraphics[width=0.9\columnwidth]{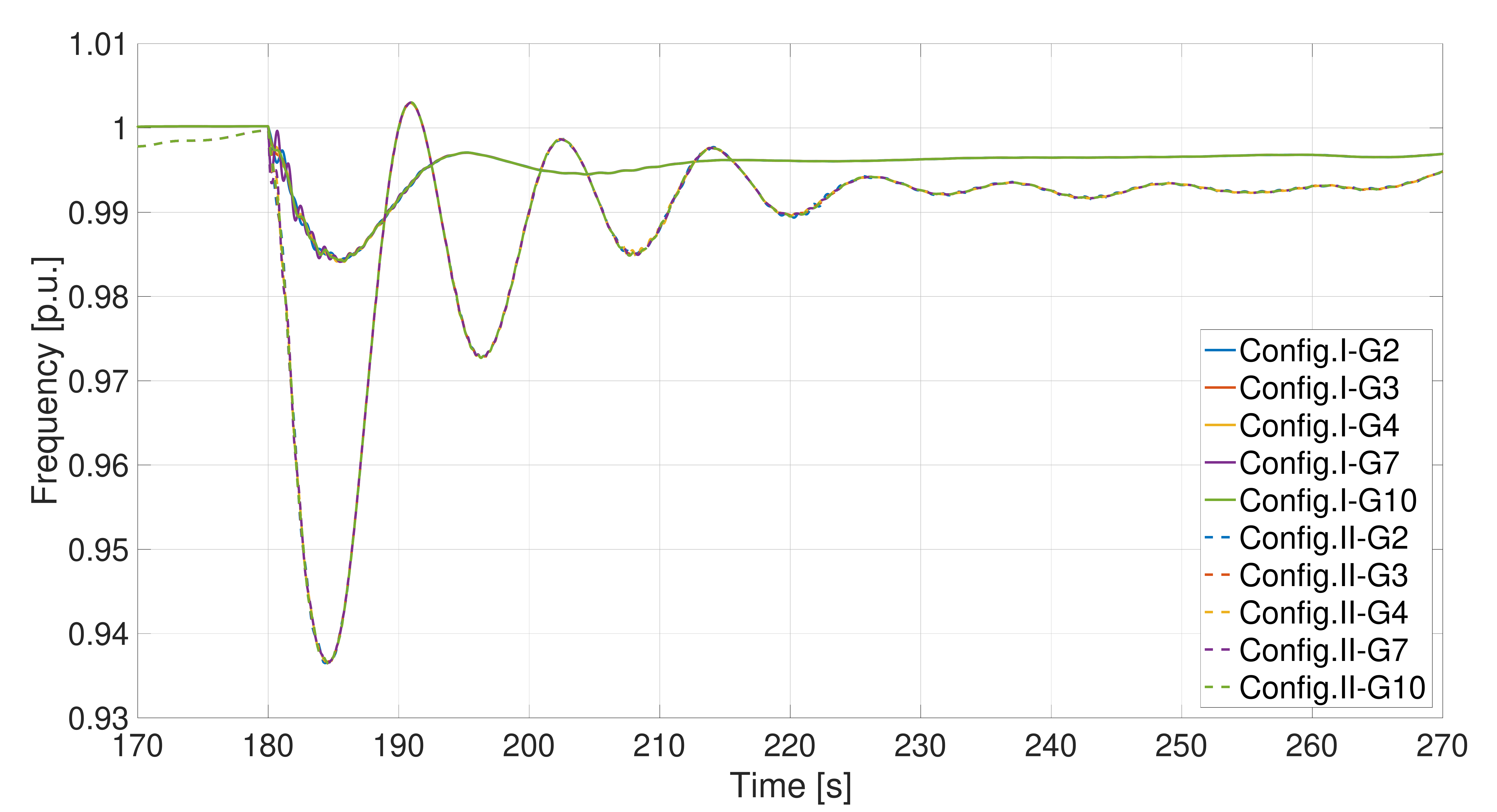}
	\caption{Frequency for Config.~I and Config.~II.}
	\label{fig:effectofreduceinertia}
\end{figure}

Fig.~\ref{fig:effectofreduceinertia} shows the system frequency for Config.~I and Config.~II. It denotes that, after the G6 tripped, the grid frequency decreases faster in Config.~II than in Config.~I. The frequency nadir for Config.~II is 0.9366 p.u. and 0.9842 p.u. for Config.~I. The frequency transient is longer in Config.~II (80~sec) than in Config.~I (30~sec). This is in-line with expectations since Config.~II has much lower system inertia than Config.~I. 

\begin{table}[t]
		\centering
		\caption{Initial Nodal Power Injections}
		\label{innodal}
		\begin{tabular}{|l|c c|c c|}
			\hline
			Unit& \multicolumn{2}{c|}{Active Power [MW]}&\multicolumn{2}{c|}{Reactive Power [MVar]}\\
			\hline
			&Config. I&Config. II &Config. I&Config. II \\
			
			G1/WP1&1353 &1335  &253 & 86\\
			G2&816 &579 &115 &56 \\
			G3& 597& 509 &70 & -61\\
			G4& 697& 545 &-56 &10 \\
			G5/WP4&406&501  &64 & 14\\
			G6&799 & 816&113 & 67\\
			G7&446 &530&-25 &-46 \\
			G8/WP2&698 & 1145 &-108 & 57\\
			G9/WP3&699 &803 &-73 &29 \\
			G10&598 &414  &41 &-87 \\			\hline
			Total& 7129&7147 &295 &206 \\
			\hline
		\end{tabular}
	\end{table}

\subsection{Compare VSC Controls in Reduced-inertia Power Grid}
\begin{figure}[ht]
	\captionsetup[subfloat]{farskip=2pt,captionskip=1pt}
	\centering
	\subfloat[Frequency.]{
		\includegraphics[width=0.9\linewidth]{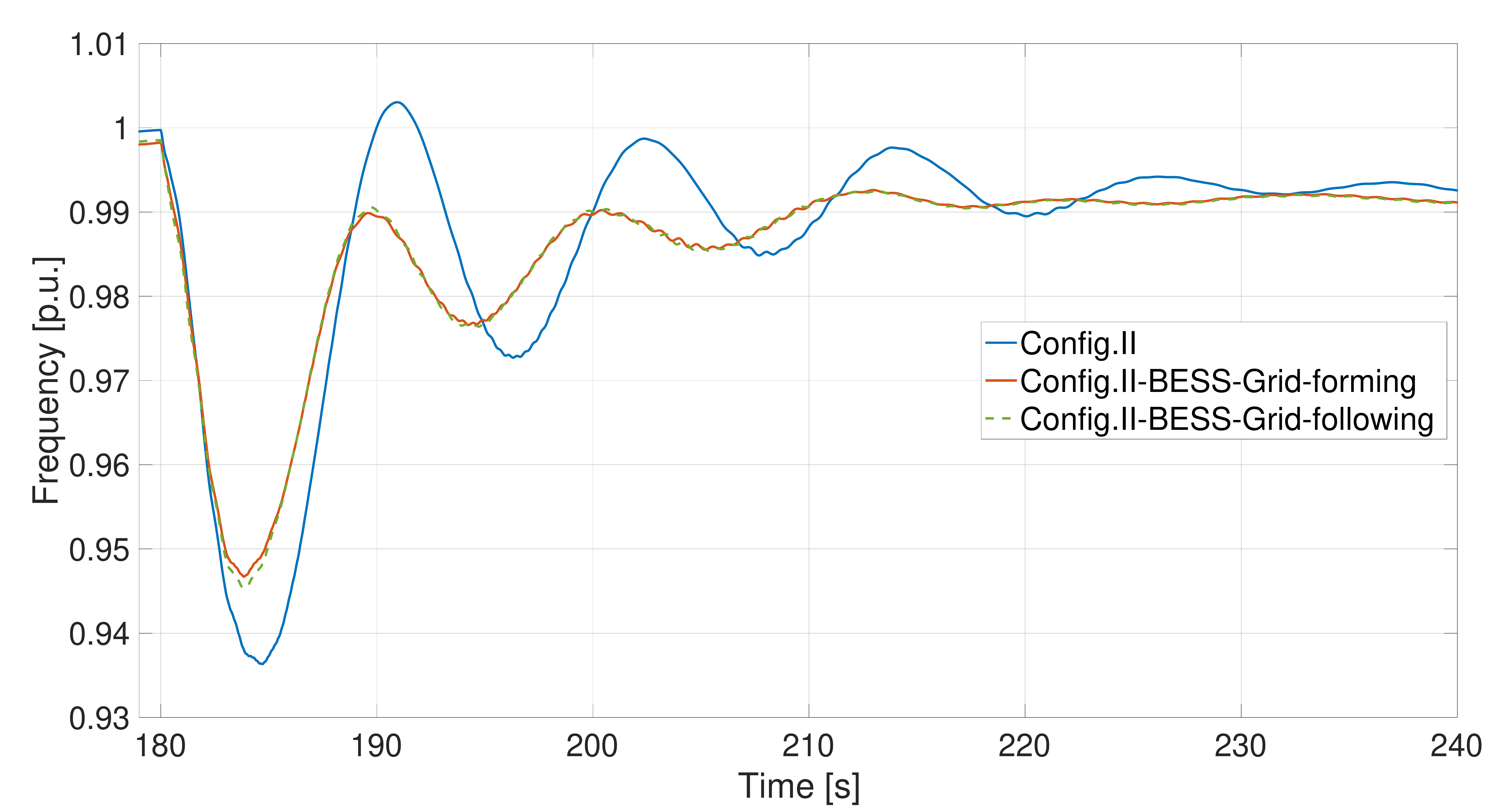}
		\label{fig:ConfigIIIf}
	}
	\hfill
	\subfloat[Active power.]{
		\includegraphics[width=0.9\linewidth]{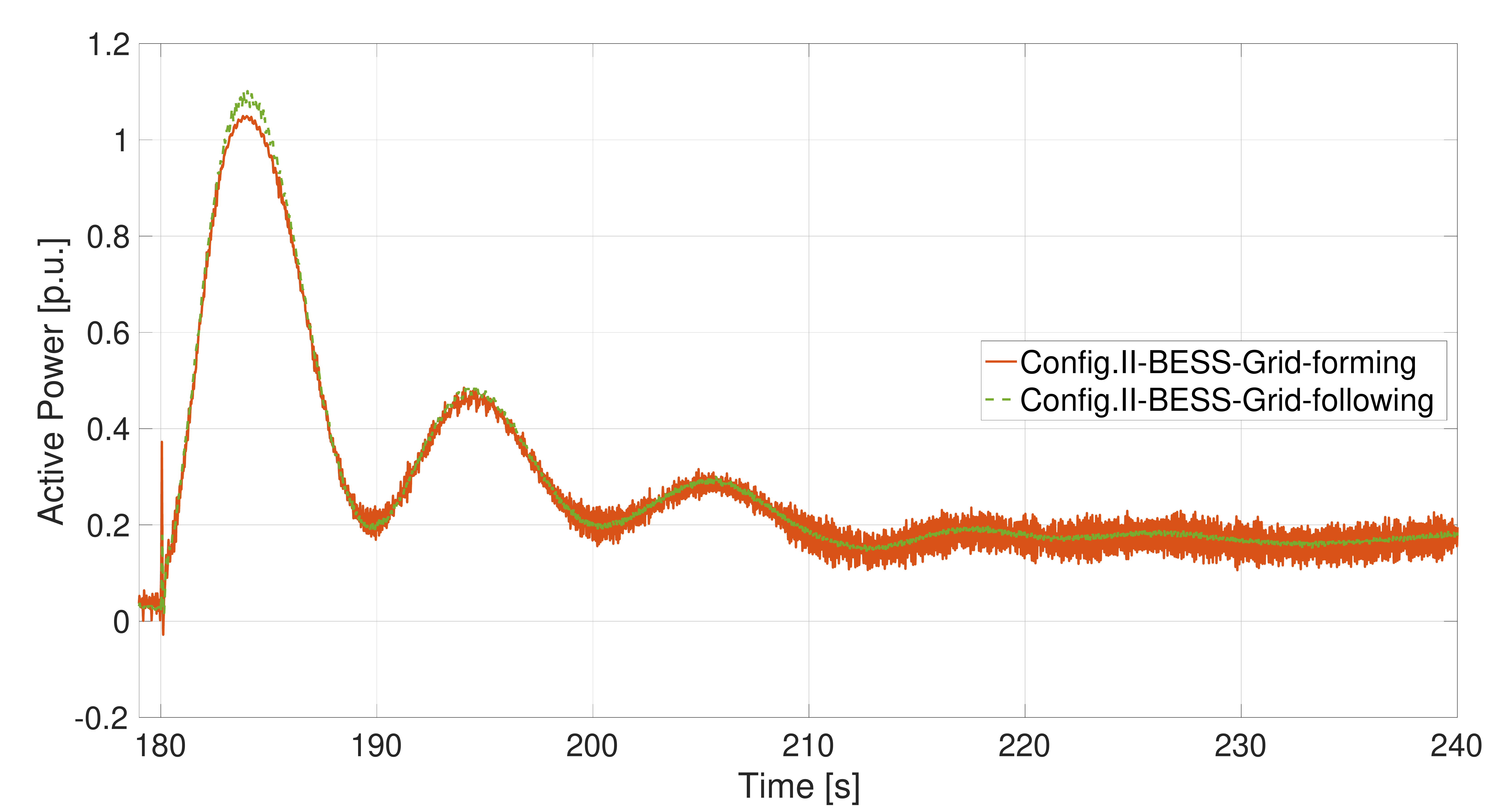}
		\label{fig:ConfigIIIP}
	}
	\hfill
	\subfloat[Reactive power.]{
		\includegraphics[width=0.9\linewidth]{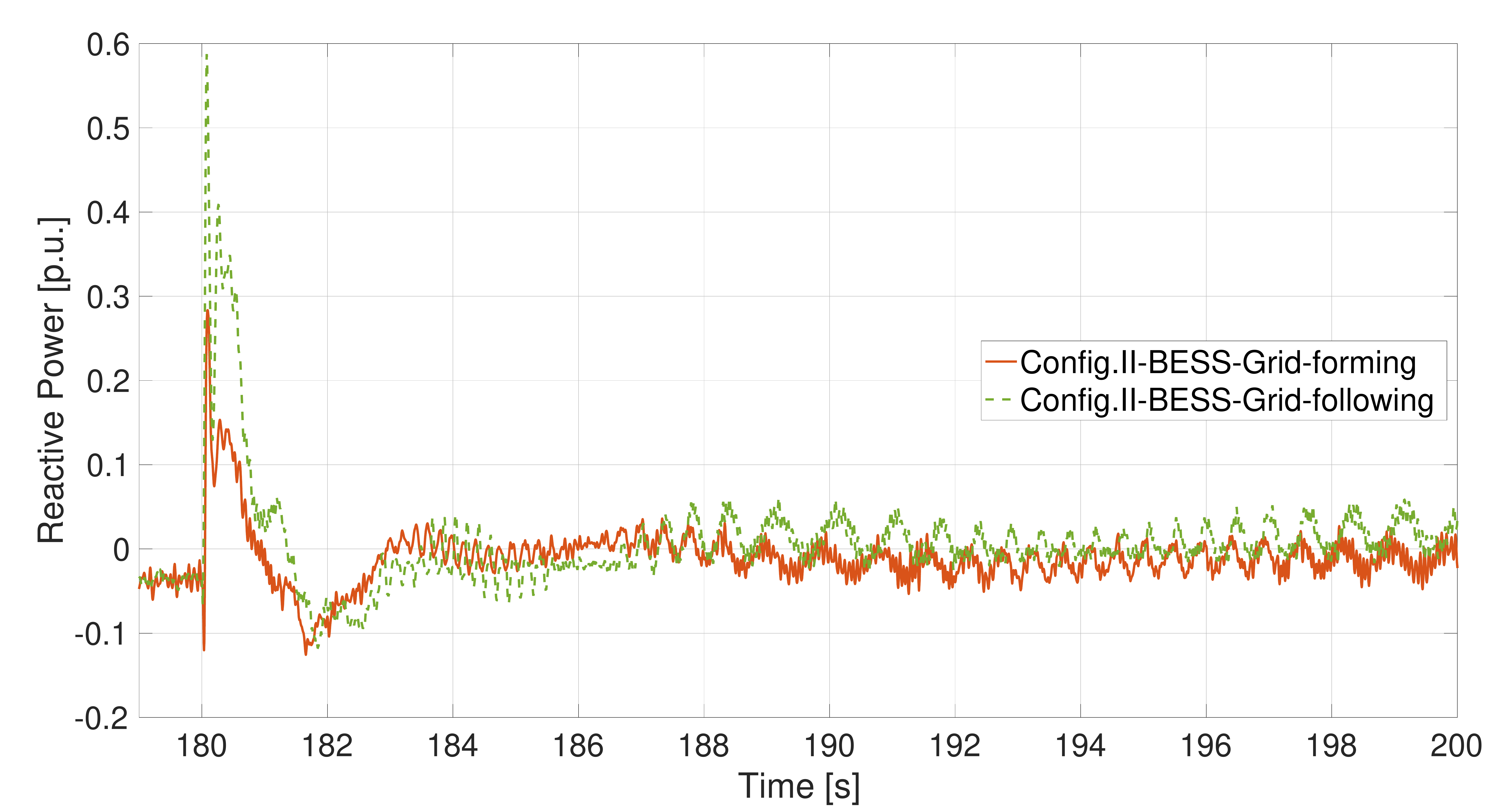}
		\label{fig:ConfigIIIQ}
	}
	\hfill
	\subfloat[Grid voltage at PCC.]{
		\includegraphics[width=0.9\linewidth]{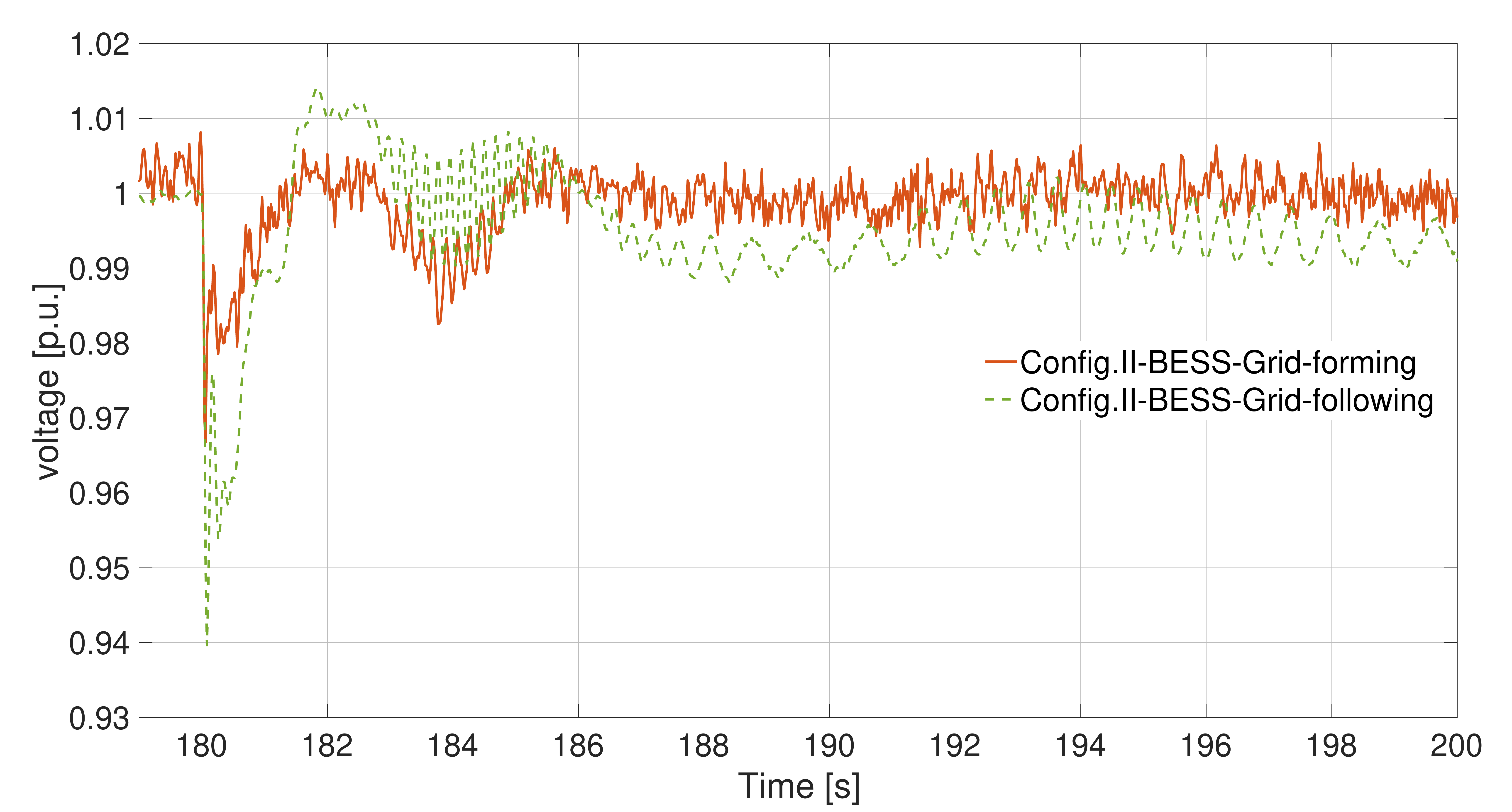}
		\label{fig:PCCvoltage}
	}
	\caption{Frequency, converter power injections and grid voltage at bus 17 in Config.~II-BESS for \textit{Case~1}.}
	\label{ConfigIII}
\end{figure}
We integrate in Config.~II a converter-interfaced BESS, modelled as described in the previous section. We denote this new configuration as Config.~II-BESS and use it to assess the performance of the grid-following and grid-forming controllers in two study cases:
\begin{itemize}
    \item \textit{Case~1}: same contingency as in the former paragraph, tripping of G6 (800~MW generation loss).
    \item \textit{Case~2}: tripping of G4 (545MW generation loss).
\end{itemize}


\begin{figure}[ht]
		\captionsetup[subfloat]{farskip=2pt,captionskip=1pt}
	\centering
	\subfloat[Frequency.]{
		\includegraphics[width=0.9\linewidth]{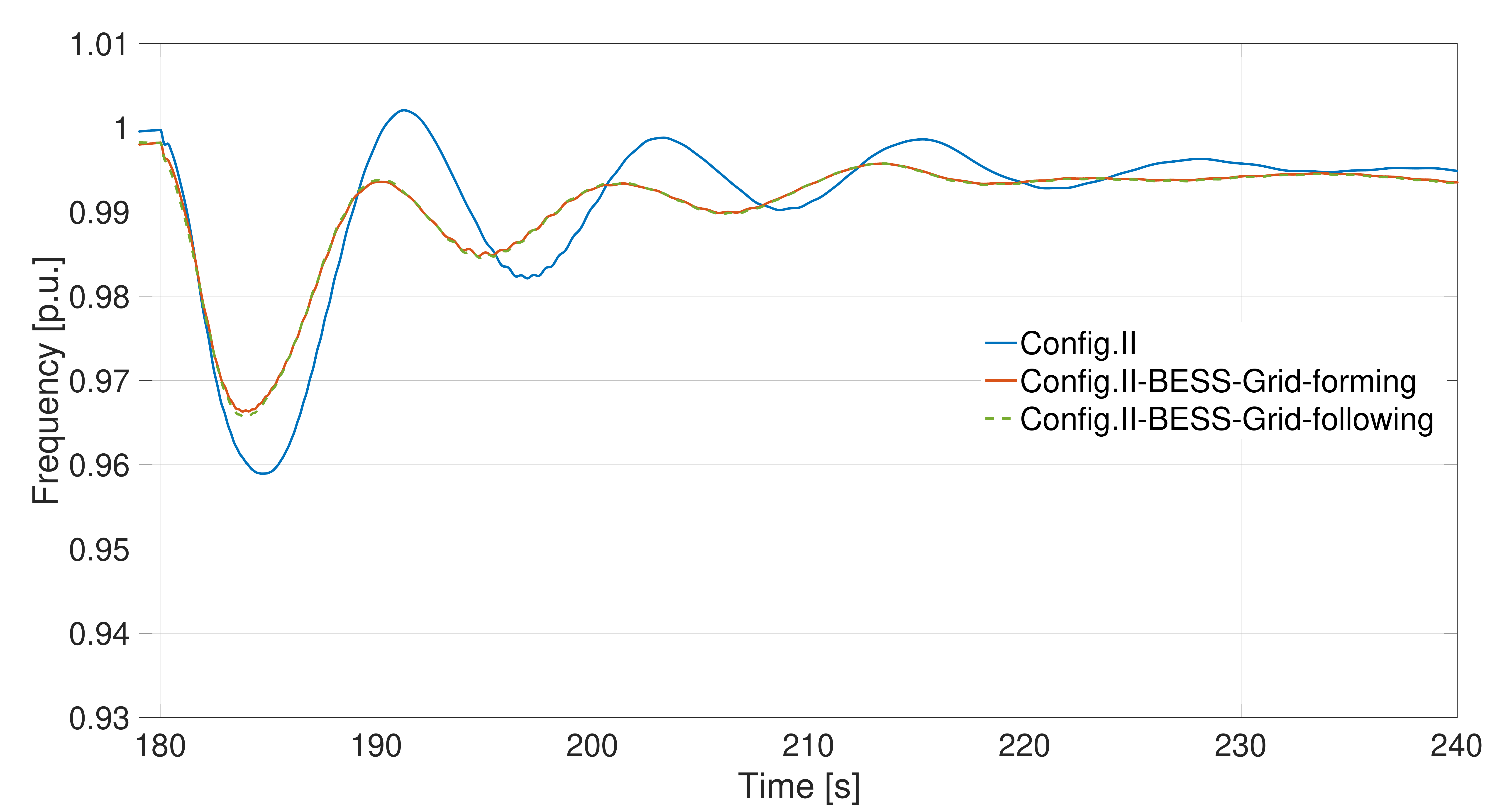}
		\label{fig:ConfigIIIfcase2}
	}
	\hfill
	\subfloat[Active power.]{
		\includegraphics[width=0.9\linewidth]{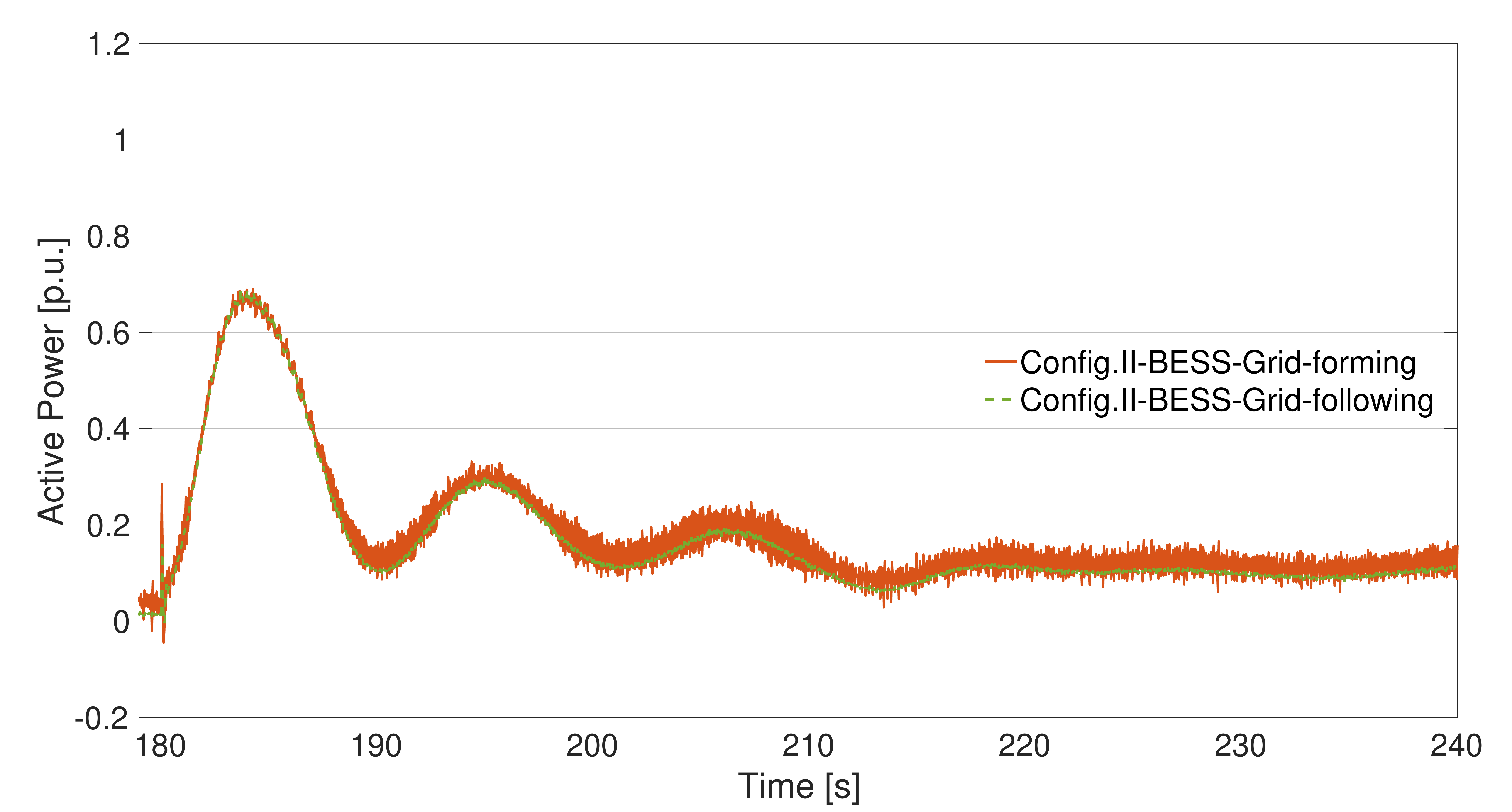}
		\label{fig:ConfigIIIPcsae2}
	}
	\hfill
	\subfloat[Reactive power.]{
		\includegraphics[width=0.9\linewidth]{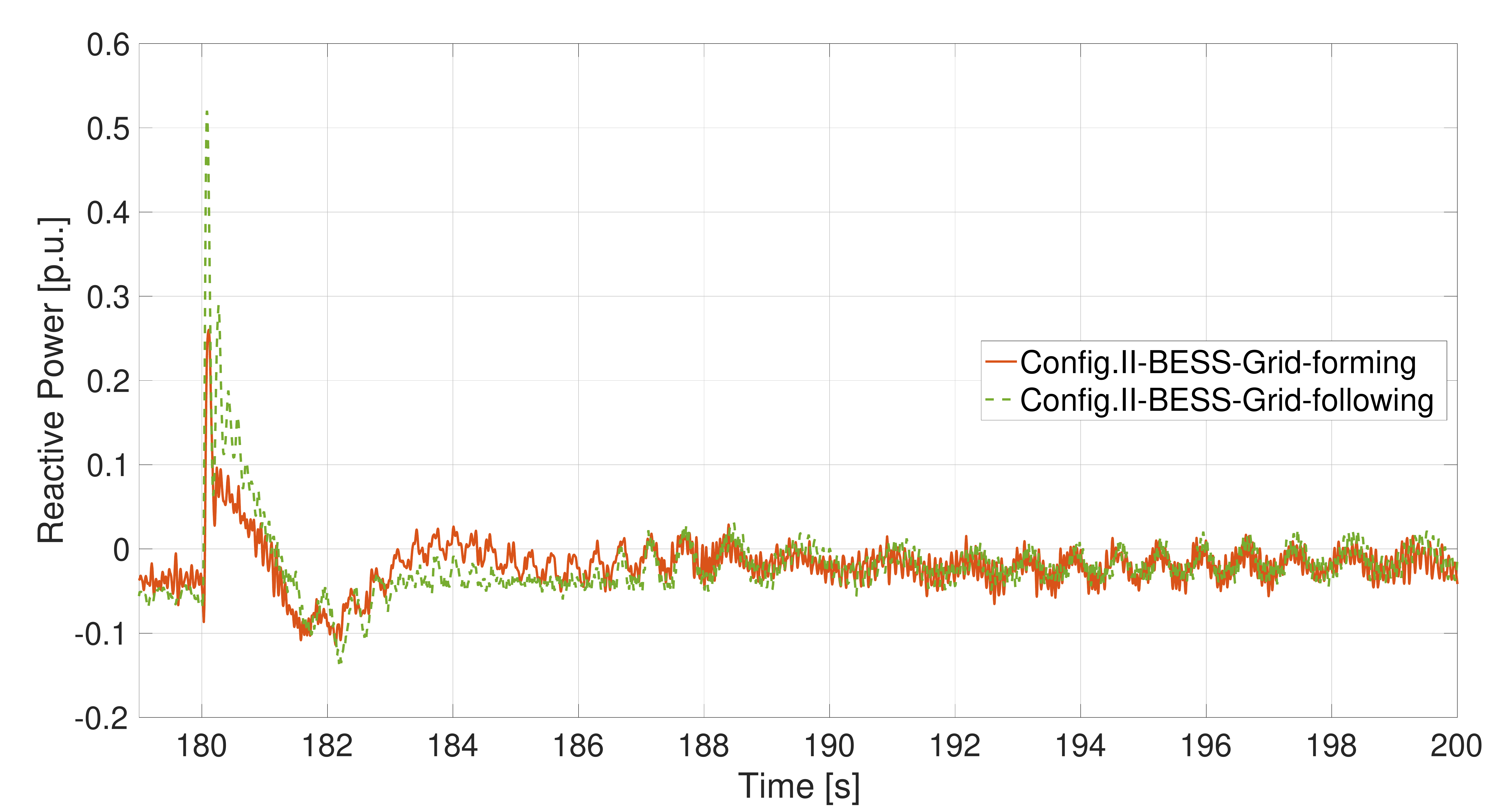}
		\label{fig:ConfigIIIQcase2}
	}
	\hfill
	\subfloat[Grid voltage at PCC.]{
		\includegraphics[width=0.9\linewidth]{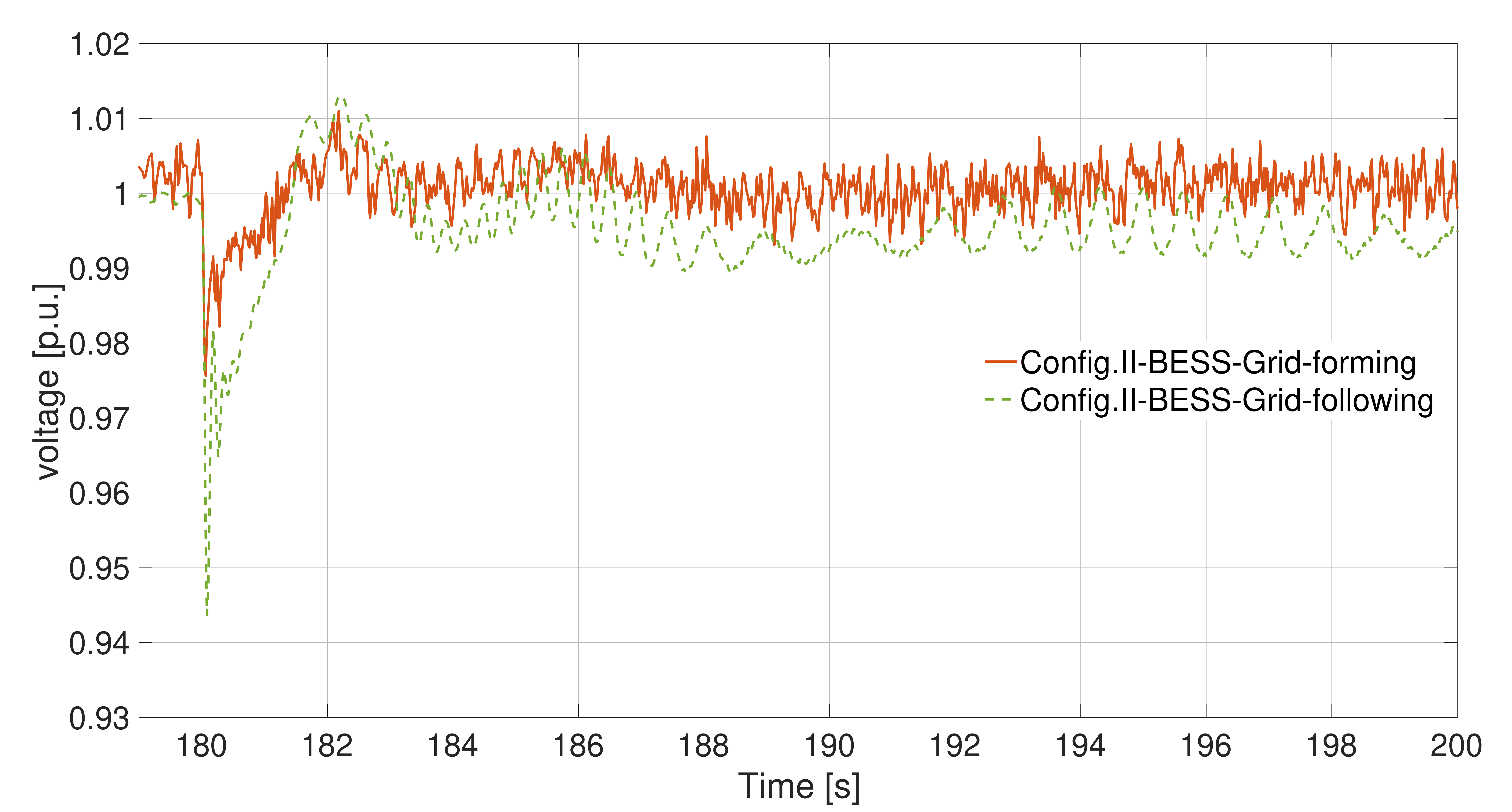}
		\label{fig:PCCvoltagecase2}
	}
	\caption{Frequency, converter power injections and grid voltage at bus 17 in Config.~II-BESS for \textit{Case~2}.}
	\label{ConfigIIIcase2}
\end{figure}
\subsubsection{Case~1}

In \textit{Case~1}, we reproduce a contingency for Config.~II-BESS the same as in Section.\ref{ConfigIvsConfigIIresult}.
Fig.~\ref{fig:ConfigIIIf} shows a comparison of the frequency behaviour in Config.~II vs Config.~II-BESS.
It shows that the VSC-based BESS achieves to increasing frequency Nadir from 0.9366 p.u. to 0.9480 p.u. and a better damping of the frequency oscillations by decreasing the overall transient interval from 80~s to 40~s. 
\begin{figure}[ht]
		\captionsetup[subfloat]{farskip=2pt,captionskip=1pt}
    \centering
    \subfloat[DC voltage.]{
    \includegraphics[width=0.9\linewidth]{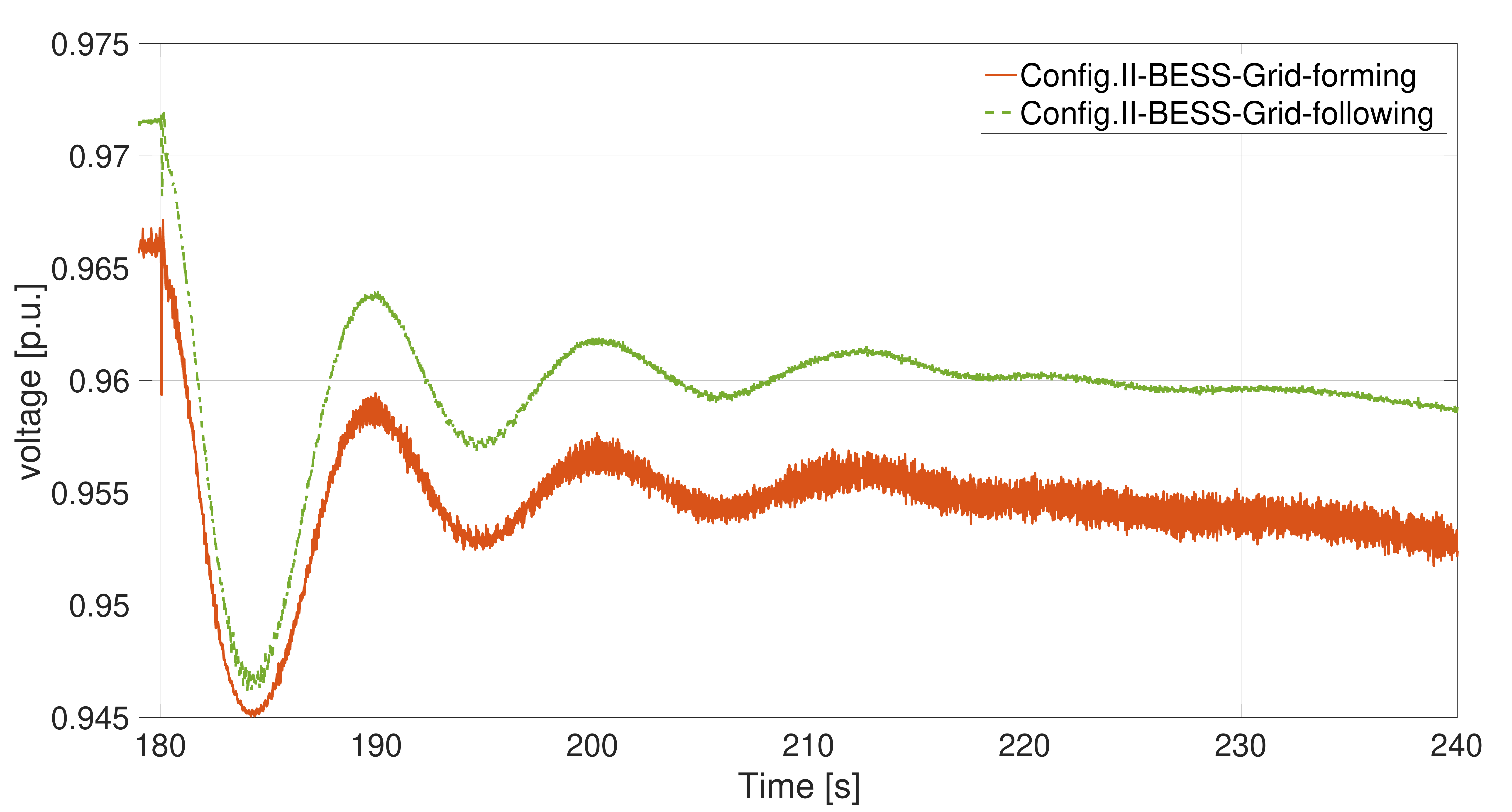}
    \label{fig:ConfigIIIvdc1}
    }
    \hfill
    \subfloat[DC current.]{
    \includegraphics[width=0.9\linewidth]{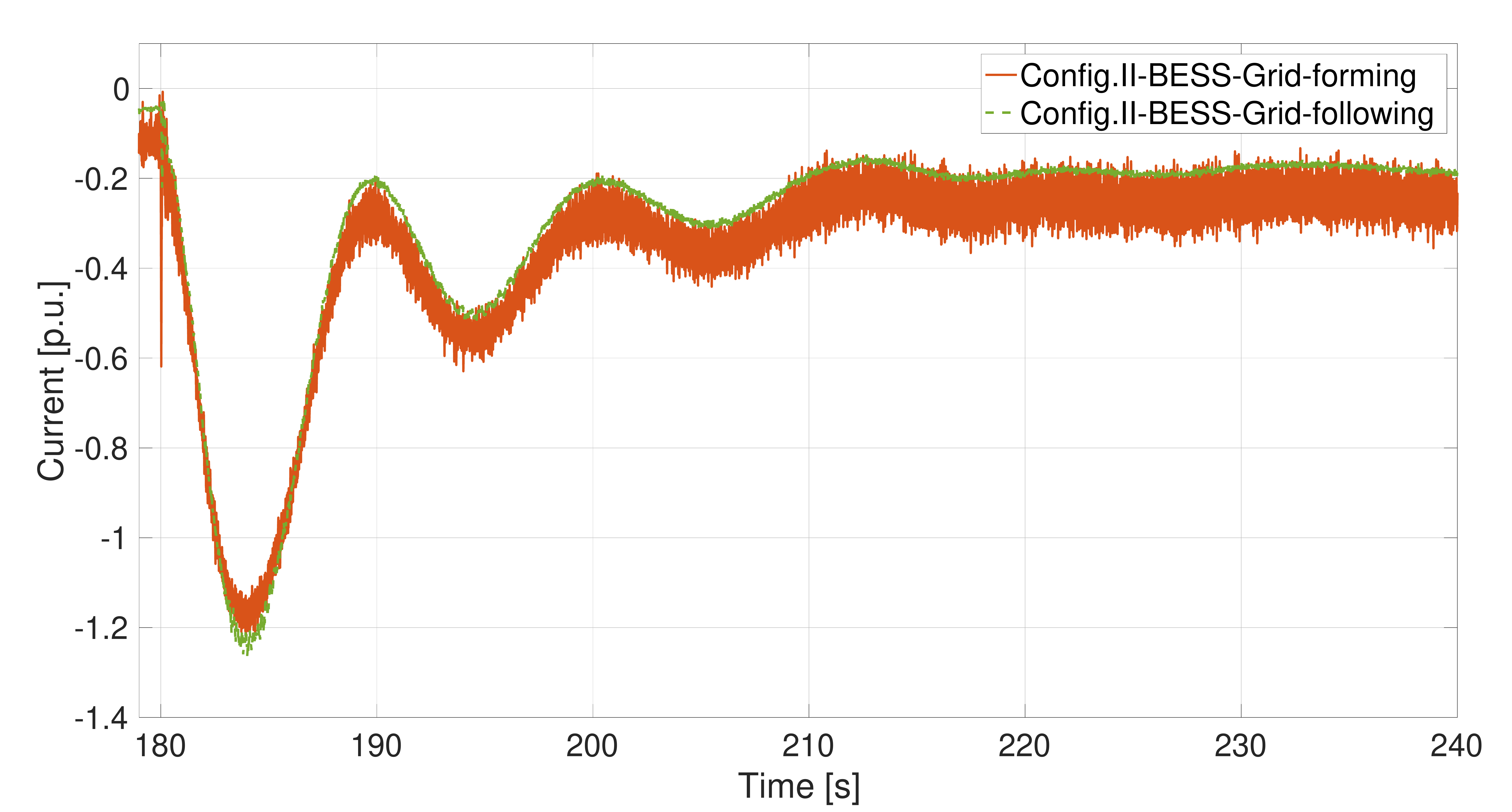}
    \label{fig:ConfigIIIIdc1}
	}
	\hfill
    \subfloat[Battery SOC.]{
    \includegraphics[width=0.9\linewidth]{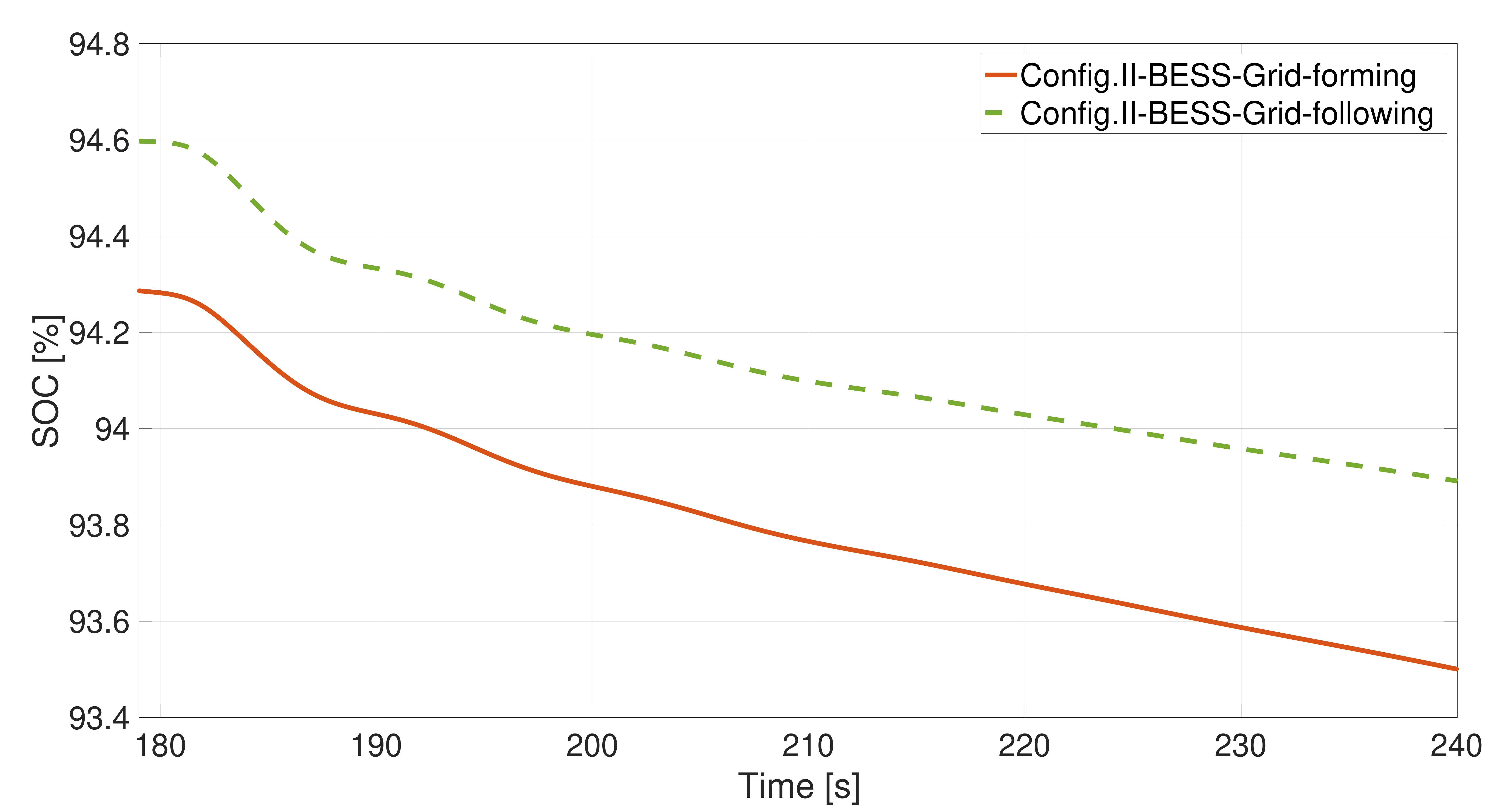}
    \label{fig:ConfigIIISOC1}
	}
    \caption{Converter DC  voltage, DC current and BESS SOC at bus 17 in Config.~II-BESS for \textit{Case~1}.}
    \label{ConfigIIIDCs1}
\end{figure}

Fig.~\ref{fig:ConfigIIIP} and Fig.~\ref{fig:ConfigIIIQ} show the active and reactive power for the installed converter unit. The grid-following and the grid-forming controllers use the same frequency droop coefficient $K_{p-f}^{following}=K_{p-f}^{forming}=20$, thus both controllers inject active power into the power system following the same droop characteristic. The considered grid-following control injects reactive power as the result of external voltage regulation, whereas the reactive power injected by the considered grid-forming control is due to the implicit coupling between active power and reactive power. As shown by Fig.~\ref{fig:ConfigIIIQ}, during the transient the reactive power injected by the grid-following converter rises up to 0.587 p.u., while the reactive power injected by the grid-forming converter only goes up to 0.284 p.u.
\begin{figure}[ht]
		\captionsetup[subfloat]{farskip=2pt,captionskip=1pt}
	\centering
	\subfloat[DC voltage.]{
		\includegraphics[width=0.9\linewidth]{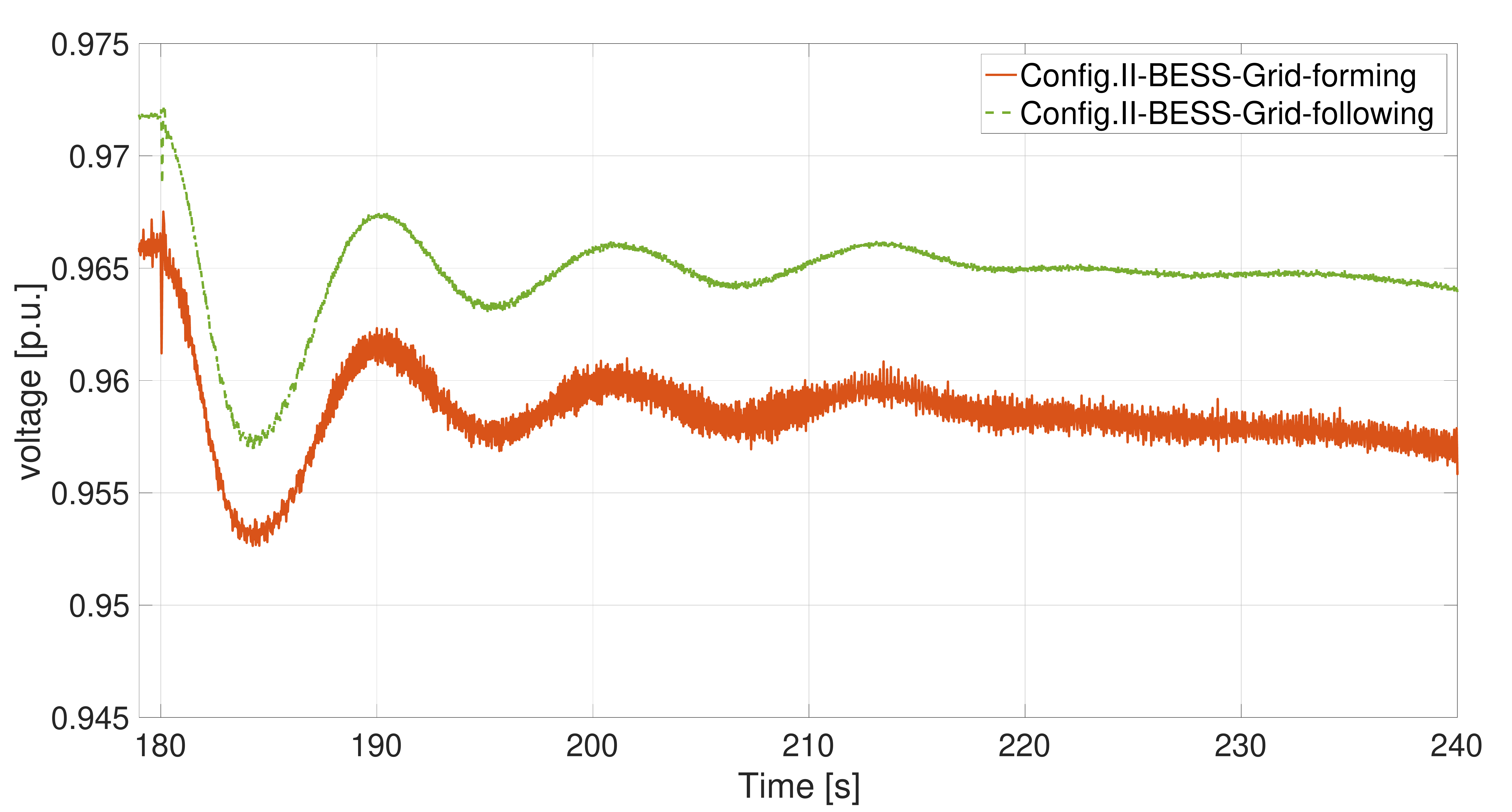}
		\label{fig:ConfigIIIvdc2}
	}
	\hfill
	\subfloat[DC current.]{
		\includegraphics[width=0.9\linewidth]{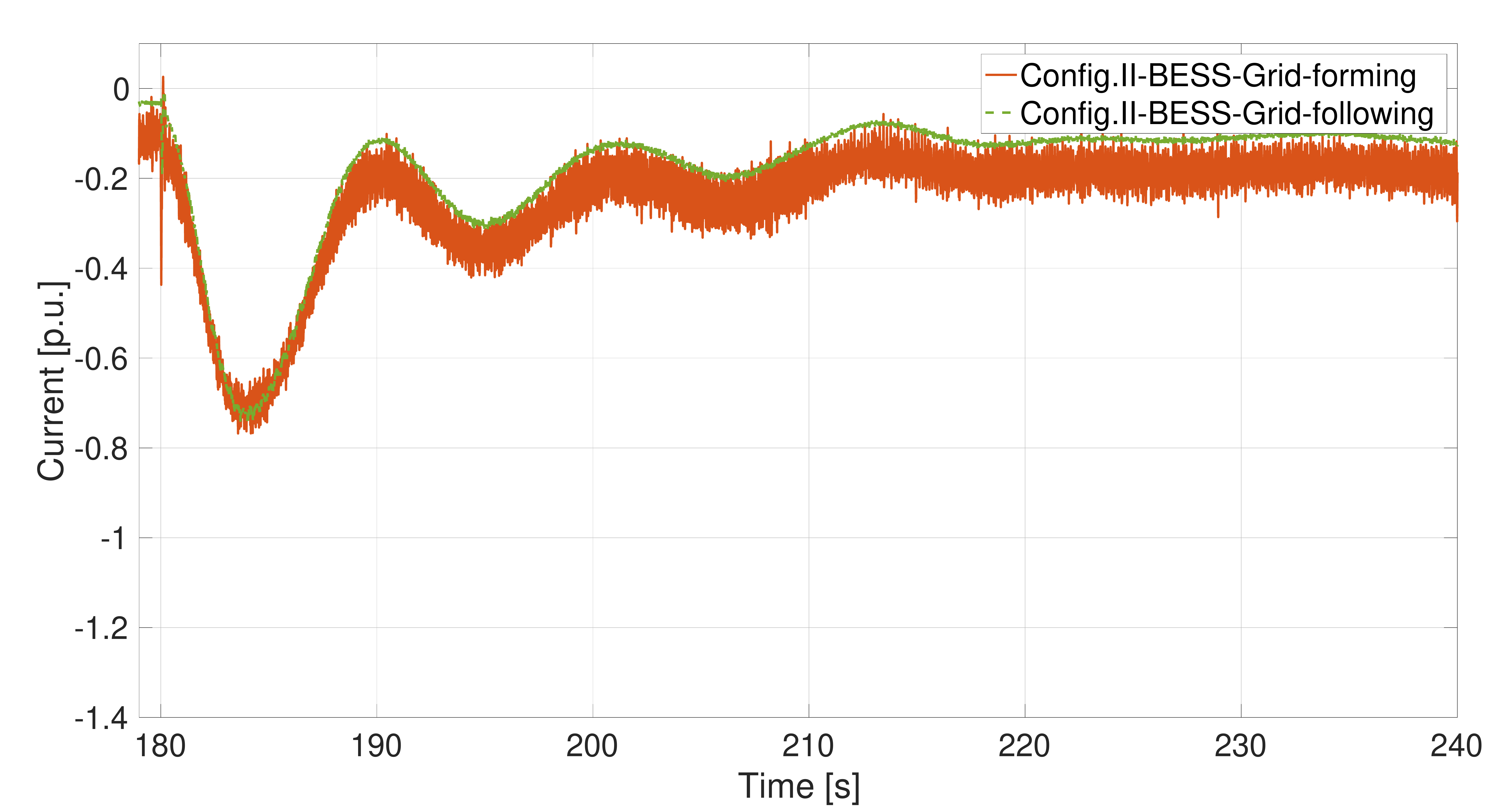}
		\label{fig:ConfigIIIIdc2}
	}
	\hfill
	\subfloat[Battery SOC.]{
		\includegraphics[width=0.9\linewidth]{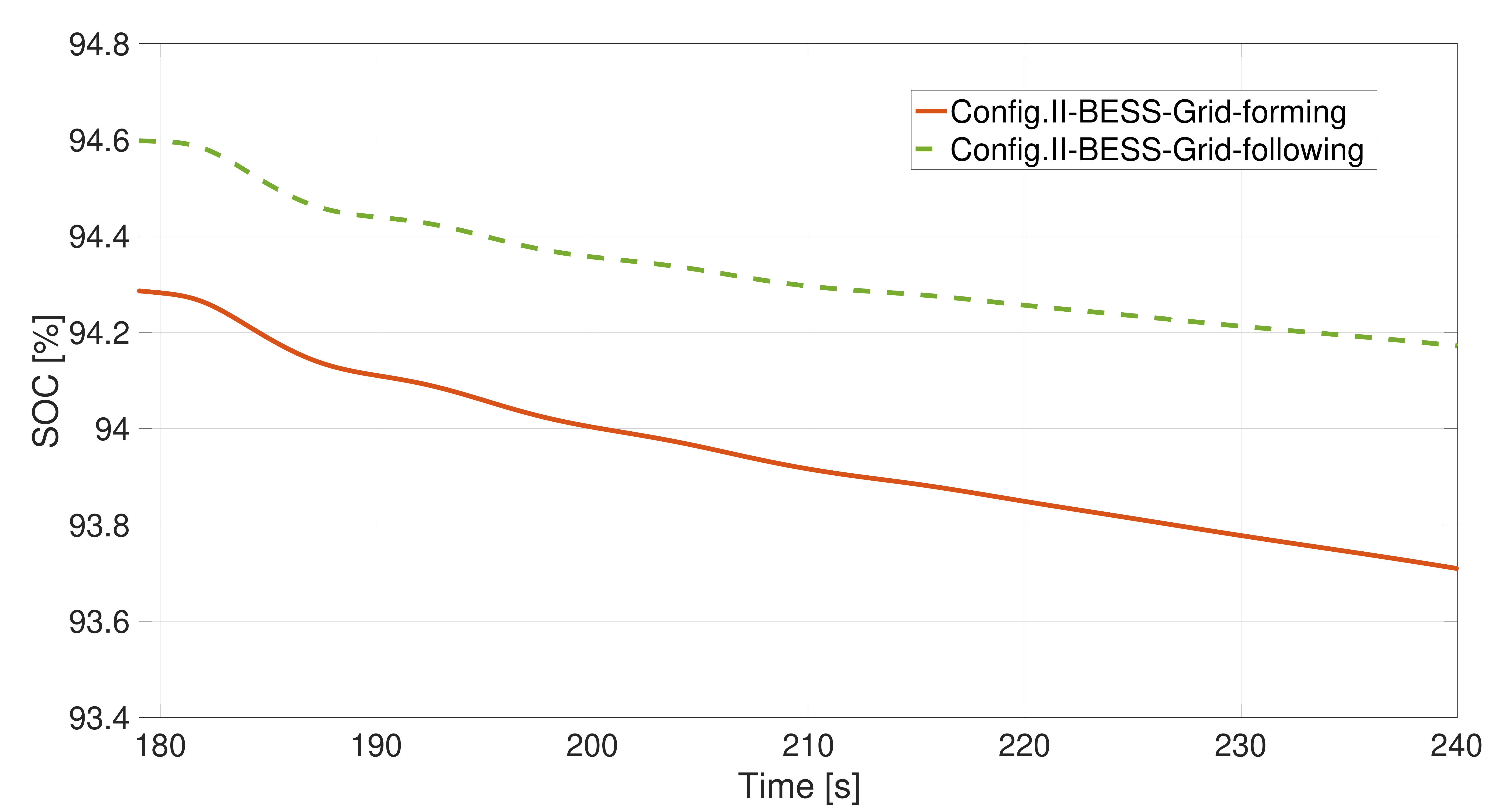}
		\label{fig:ConfigIIISOC2}
	}
	\caption{Converter DC  voltage, DC current and BESS SOC at bus 17 in Config.~II-BESS for \textit{Case~2}.}
	\label{ConfigIIIDCs2}
\end{figure}

Fig.~\ref{fig:PCCvoltage} presents the amplitudes of the grid voltage at the PCC of the installed converter unit (i.e., bus 17). It denotes that, after the contingency there is a voltage sag (i.e., decrease of 6\% of nominal voltage) within 100~ms for the grid-following converter whereas the PCC voltage for the grid-forming converter experiences a way low drop (it varies only of $\pm 3\%$). In addition, during the whole transient period, the voltage variation for the grid-following converter appears larger than for the case of the grid-forming converter. Although the grid-following control injects higher amount of reactive power, the voltage regulation result is not as good as for the grid-forming control.
This is because the grid-forming control allows the converter operating as voltage source which is capable of better sustaining the PCC voltage. 

Fig.~\ref{ConfigIIIDCs1} shows converter DC voltage, DC current, and battery SOC for \textit{Case 1}. The DC voltage varies corresponding to the change of DC current. The SOC of BESS is decreasing in a way that corresponds to the integrated of injected power into the grid due to the frequency regulation.

\subsubsection{Case~2}

To represent a less extreme contingency, in \textit{Case~2} we trip G4 to cause less generation loss. Fig.~\ref{ConfigIIIcase2} shows the simulation results of reproducing the same contingency for Config.~II and Config.~II-BESS. 

Fig.~\ref{fig:ConfigIIIfcase2} presents the frequency responses for Config.~II and Config.~II-BESS. It illustrates that the converter unit increases the frequency Nadir from 0.9589 for Config.~II to 0.9665 for Config.~II-BESS and ameliorate the frequency oscillations by decreasing the transient duration from 75~s to 35~s.

Fig.~\ref{fig:ConfigIIIPcsae2} and Fig.~\ref{fig:ConfigIIIQcase2} show the active and reactive power injected by the converter unit. For both the grid-following and grid-forming control, the injected active power tracks frequency deviations accordingly with their droop coefficients. During the transient, the reactive power injected by the grid-following converter rises up to 0.520 p.u., while the reactive power injected by the grid-forming converter only goes up to 0.260 p.u.

Fig.~\ref{fig:PCCvoltagecase2} shows the amplitude of the grid voltage at the PCC of the converter units. It demonstrates the benefit of the grid-forming converter as voltage source in preventing the PCC voltage from large variation. In contrast, the grid-following converter experiences a voltage sag ($-6\%$ of nominal voltage) within 100~ms after the contingency and a generally higher voltage variation during the transient.

Fig.~\ref{fig:ConfigIIISOC2} shows converter DC voltage, DC current, and battery SOC for \textit{Case 2}. As for the previous \textit{Case 1}, the DC voltage varies as a function of DC current and the SOC of BESS decreases as a result of the BESS frequency regulating action.

\section{Conclusions}
In this paper we investigated the impact of VSCs on the dynamics of a reduced-inertia grid that interfaces a mix of synchronous machines and power-electronics-interfaced wind turbines. To this end, we proposed three 39-bus power system configurations as an extension of the IEEE 39-bus benchmark power system.
The first one corresponds to the original benchmark network. The second configuration replaces four synchronous machine-based power plants with type-3 wind turbines. The third configuration is identical to the second with the exception of including a power electronic-interfaced BESS. 
Correspondingly, we built three full-replica dynamic models that are executed on a real-time simulator to reproduce the same contingencies and conduct post-contingency analysis with respect to the system dynamics.

The simulation results verified the substantial influence of inertia reduction on the post-contingency dynamics of the power system and quantitatively proved that the connected VSC, implemented with the grid-following control with supporting mode or the PLL-free grid-forming control, can assist in limiting the frequency decreasing and in damping the frequency oscillations. The performance of the grid voltages at the PCC of the converter has demonstrated the benefit of the grid-forming converter to maintain the PCC voltage during transient, along with an important improvement of the post-contingency frequency transient in terms of both Nadir and damping.

Our future work will utilize long time steady-state simulations to quantify and analyze the benefits of BESSs for frequency response services.

\IEEEtriggercmd{\enlargethispage{-5in}}


%

\bibliographystyle{IEEEtran}
\bibliography{biblio.bib}

\end{document}